\documentclass[11pt,a4paper]{article}
\usepackage[utf8]{inputenc}
\usepackage{graphicx,verbatim,array,multicol,courier}
\usepackage{amsmath,amssymb,amsthm,mathrsfs, mathtools}
\usepackage{color, graphics, graphicx, siunitx}
\usepackage{amsfonts, dsfont, bm}
\usepackage{natbib} 
\usepackage{geometry}
\usepackage{enumitem,float}
\usepackage{lscape}
\usepackage[pdftex,bookmarks,colorlinks]{hyperref}
\usepackage[dvipsnames]{xcolor}
\usepackage{tabularx}
\usepackage[toc,title,page]{appendix}
\usepackage{siunitx}
\usepackage[labelformat=simple]{subfig}
\usepackage[linesnumbered,ruled,vlined]{algorithm2e}
\usepackage{booktabs}

\newlist{inparaenum}{enumerate}{2}
\setlist[inparaenum,1]{label=(\alph*)}
\setlist[inparaenum,2]{label=(\roman{inparaenumi}\emph{\alph*})}

\makeatletter
\def\adl@drawiv#1#2#3{%
        \hskip.5\tabcolsep
        \xleaders#3{#2.5\@tempdimb #1{1}#2.5\@tempdimb}%
                #2\z@ plus1fil minus1fil\relax
        \hskip.5\tabcolsep}
\newcommand{\cdashlinelr}[1]{%
  \noalign{\vskip\aboverulesep
           \global\let\@dashdrawstore\adl@draw
           \global\let\adl@draw\adl@drawiv}
  \cdashline{#1}
  \noalign{\global\let\adl@draw\@dashdrawstore
           \vskip\belowrulesep}}
\makeatother

\numberwithin{equation}{section}
\theoremstyle{definition}
\newtheorem{defi}{Definition}[section]

\theoremstyle{plain}
\newtheorem{theo}[defi]{Theorem}
\newtheorem{prop}[defi]{Proposition}

\newtheorem{cor}[defi]{Corollary}

\theoremstyle{remark}
\newtheorem{rem}[defi]{Remark}

\theoremstyle{example}
\newtheorem{ex}[defi]{Example}

\newcommand{\diff}{\mathrm{d}}

\definecolor{navy}{rgb}{0,0,0.502}
\definecolor{brown}{rgb}{0.59, 0.29, 0.0}
\def\indic{\mathds{1}}

\hypersetup{colorlinks,%
	citecolor=blue,%
	filecolor=green,%
	linkcolor=red,%
	urlcolor=violet,%
}

\newcommand{\bbP}{\mathbb{P}}

\newcommand{\Real}{\mathbb{R}}

\newcommand{\DoA}{\mathcal{D}}
\newcommand{\Prob}{\mathbb{P}}

\newcommand{\bftheta}{{\boldsymbol{\theta}}}
\newcommand{\bfxi}{{\boldsymbol{\xi}}}
\newcommand{\bfvartheta}{{\boldsymbol{\vartheta}}}

\newcommand{\bfmu}{{\boldsymbol{\mu}}}
\newcommand{\bfbeta}{{\boldsymbol{\beta}}}

\newcommand{\bfSigma}{{\boldsymbol{\Sigma}}}
\newcommand{\bfOmega}{{\boldsymbol{\Omega}}}

\newcommand{\bfB}{{\boldsymbol{B}}}
\newcommand{\bfC}{{\boldsymbol{C}}}
\newcommand{\bfJ}{{\boldsymbol{J}}}

\newcommand{\bfA}{{\boldsymbol{A}}}
\newcommand{\bfD}{{\boldsymbol{D}}}
\newcommand{\bfE}{{\boldsymbol{E}}}
\newcommand{\bfX}{{\boldsymbol{X}}}
\newcommand{\bfY}{{\boldsymbol{Y}}}
\newcommand{\bfM}{{\boldsymbol{M}}}
\newcommand{\bfS}{{\boldsymbol{S}}}
\newcommand{\bfV}{{\boldsymbol{V}}}
\newcommand{\bfZ}{{\boldsymbol{Z}}}

\newcommand{\bfI}{{\boldsymbol{I}}}

\newcommand{\bfzero}{{\boldsymbol{0}}}
\newcommand{\bfone}{{\boldsymbol{1}}}

\DeclareMathOperator*{\argmax}{arg\,max}

\title{Accurate Bayesian inference for tail risk extrapolation in time series}
\author{David L. Carl, Simone A. Padoan and Stefano Rizzelli}
\begin{document}
\maketitle
\begin{abstract}
Accurately quantifying tail risks—rare but high-impact events such as financial crashes or extreme weather—is a central challenge in risk management, with serially dependent data. We develop a Bayesian framework based on the Generalized Pareto (GP) distribution for modeling threshold exceedances, providing posterior distributions for the GP parameters and tail quantiles in time series. Two cases are considered: extrapolation of tail quantiles for the stationary marginal distribution under $\beta$-mixing dependence, and dynamic, past-conditional tail quantiles in heteroscedastic regression models. The proposal yields asymptotically honest credible regions, whose coverage probabilities converge to their nominal levels.
We establish the asymptotic theory for the Bayesian procedure, deriving conditions on the prior distributions under which the posterior satisfies key asymptotic properties. To achieve this, we first develop a likelihood theory under serial dependence, providing local and global bounds for the empirical log-likelihood process of the misspecified GP model and deriving corresponding asymptotic properties of the Maximum Likelihood Estimator (MLE). Simulations demonstrate that our Bayesian credible regions outperform naïve Bayesian and MLE-based confidence regions across several standard time-series models, including ARMA, GARCH, and Markovian copula models. Two real-data applications—to U.S. interest rates and Swiss electricity demand—highlight the relevance of the proposed methodology.
\end{abstract}

%
\section{Introduction}\label{se:intro}
%
In many disciplines—ranging from economics and finance to the environmental and applied sciences—there is strong interest in analyzing temporally dependent data, such as financial returns, air pollution levels, electricity prices or consumption, with the aim of understanding and predicting future behavior. Accounting for serial dependence is crucial in such analyses, as neglecting it can lead to misleading estimates. A key challenge arises when extrapolating far into the tail of the underlying distribution—beyond the most extreme observed values—to estimate the probability of events more severe than any previously recorded. These low-probability, high-impact events, commonly referred to as ``tail risks'' (e.g., financial crises, extreme weather events), can profoundly disrupt society.  Their extrapolation, coupled with accurate uncertainty quantification, is a central goal in risk management, underpinning both adaptation and mitigation strategies. In particular, quantifying the variability of inferential procedure in the extrapolation regime for time series remains a significant challenge \citep[e.g.][]{bucher2025bootstrapping, de2024bootstrapping, nolde2021extreme, bucher2021horse, de2016adapting}.

To address this issue, we develop a Bayesian framework that delivers posterior distributions for extrapolating tail risks in two key settings. The first concerns the extrapolation of tail quantiles of the stationary marginal distribution of a time series, under the $\beta$-mixing condition. This condition encompasses many standard models in finance, including ARMA, nonlinear autoregressive processes, and GARCH.
The second setting focuses on the extrapolation of dynamic tail quantiles—conditional on the past—in heteroscedastic regression models with strictly stationary error sequences. This class includes models with independent innovations, such as ARMA and GARCH, as well as models with dependent innovations, such as ARMA–GARCH, discrete-time diffusion models, and more general CARMA processes. Our framework builds upon the modeling of suitably normalized threshold exceedances using the Generalized Pareto (GP) distribution (\citealp[e.g.][Ch. 4]{coles2001}, \citealp{balkema1974}).

For independent observations, there exists an extensive literature on Bayesian approaches to threshold-exceedance-based inference (see, e.g., \citealp{de2022extreme, northrop2016, tancredi2006accounting, coles2003anticipating, coles1996bayesian}). However, methods offering rigorous asymptotic guarantees have only recently begun to emerge (e.g., \citealt{dombry2023}; see also \citealp{padoan2024, padoan2022consistency} for the block maxima case). The difficulty stems from the intricate inferential theory of the GP distribution \citep[e.g.][]{haan2006}. The GP model is inherently misspecified for threshold exceedances, since in practice the threshold must be fixed. Moreover, the GP family is {\it irregular}: the sign of the shape parameter determines the support of the distribution, and the scale parameter depends on the chosen threshold rather than being fixed, as in more conventional models \citep[e.g.][]{vaart1998}. These features make likelihood-based inference—including Bayesian approaches—particularly challenging.

In the time series setting, inference becomes even more challenging. In this article, we show in-fact that inference based on the asymptotic normality of the Maximum Likelihood Estimator (MLE), obtained from an independence GP likelihood, remains valuable: it provides confidence regions for the quantities of interest (the GP parameters and tail quantiles) that are {\it asymptotically honest}—that is, their coverage probability converges the nominal level for increasing sample size (in a bias-free setting). In contrast, the corresponding Bayesian inference does not share this property.

To the best of our knowledge, GP-based Bayesian inference for extrapolating tail risks in time series settings remains largely unexplored. The main contributions of this work are as follows. We propose a general Bayesian framework for constructing posterior distributions—both for the GP parameters and for tail quantiles—in the two settings discussed above: (i) extrapolation of tail quantiles for the stationary marginal distribution, and (ii) inference on dynamic tail quantiles. The proposed approach yields credible regions that are asymptotically honest, ensuring that their coverage probabilities converge to the nominal level.

To achieve this goal, we pursued two main directions.
First, we developed a comprehensive likelihood framework under serial dependence to establish the asymptotic theory for the proposed Bayesian procedure. Specifically, we derived local and global bounds for the empirical log-likelihood process of the misspecified GP model, together with a local asymptotically Gaussian expansion (detailed in the Supplement), which in turn provides contraction rates for the MLE of the GP parameters. A byproduct of these results is that the MLE, computed over the full parameter space, is consistent and uniquely maximizes the likelihood in the data-dependent setting—an open question previously unresolved and extending the results of \citet{dombry2023} in the independence case. Furthermore, as a key step toward implementing confidence regions based on the MLE’s asymptotic normality and constructing our Bayesian procedure, we propose a consistent estimator of the MLE asymptotic variance–covariance matrix.

Second, we developed the asymptotic theory for our Bayesian approach. We establish general and practical conditions on the prior distributions for the GP parameters under which the corresponding posterior distribution satisfies key asymptotic properties—namely, consistency with a contraction rate and the Bernstein–von Mises (BvM) theorem. Our framework accommodates a broad class of priors, including informative proper priors, informative empirical Bayes priors, and vague empirical Bayes priors. Although a version of the BvM theorem for misspecified models exists in the independent case \citep{kleijn2012bernstein}, it does not extend to our time-series setting, and the GP class fails to satisfy its required smoothness and regularity conditions. Finally, we introduce a new construction that derives, from the posterior of the GP parameters, the posterior distribution of tail quantiles, which retains the same theoretical guarantees.

We road-test the proposed Bayesian framework through an extensive simulation study, demonstrating the superior accuracy of our credible regions compared with those obtained using a naïve Bayesian approach and MLE-based confidence regions. The evaluation covers several widely used time series models, including ARMA, ARCH, and Markovian copula models (for estimating GP parameters and tail quantiles of the stationary marginal distribution), as well as ARMA and ARMA–GARCH models (for estimating GP parameters and dynamic, past-conditional tail quantiles).
We further illustrate the practical relevance of our approach through two empirical applications. The first analyzes U.S. one-month interest rates, whose dynamics are typically modeled by continuous-time diffusion processes observed at discrete intervals (e.g., CARMA models). The second examines electricity consumption in Switzerland. In both cases, forecasting future tail risks is of particular practical importance. 

Our methodology is implemented in the {\bf R} package {\tt ExtremeRisks}, available on CRAN. Section \ref{eq:backg} introduces the statistical framework and presents preliminary results on likelihood-based inference. Section \ref{sec:static_inference} describes our Bayesian approach and establishes its theoretical properties. The methodology is then applied to simulated data in Section \ref{sec:simulations} and to empirical analyses of interest rates and electricity demand in Section \ref{sec:realanalysis}. The online supplementary material provides the technical conditions, additional simulation results, further details on the empirical analyses, and all proofs.

%
%
\section{Background}\label{eq:backg}
%
%
%
%
\subsection{Statistical model and time series setting}\label{eq:stat_mod}
%
%

We consider a (strictly) stationary time series $(X_i)_{i\geq 1}$ with stationary cumulative distribution function (cdf) $F$, whose right-end point is $x_E=\sup (x: F(x)< 1)$. Assume that $F$ is an unknown continuous distribution belonging to the domain of attraction of a Generalised Extreme Value (GEV) distribution, denoted by  $F\in\mathcal{D}(G_\gamma)$, with $\gamma\in\Real$ (e.g., \citealp[][Ch. 8]{falk2010laws}, \citealp[][Ch. 8]{embrechts1997modelling}, \citealp[][Ch. 3]{leadbetter1983}).  This condition implies that for $m=1,2,\ldots$, there exist normalising sequences $a(m)>0$ and $b(m)\in\Real$, such that $F^m(a(m)z+b(m))\to G_\gamma(z)$, as $m\to\infty$, where $G_\gamma$ is the GEV distribution \citep[e.g.][]{haan2006, coles2001}. 
The parameter $\gamma\in\Real$, known as the {\it extreme value index}, 
characterises the tail heaviness of $F$, which is: short-, light- or heavy-tailed, depending on whether $\gamma<0$, $\gamma=0$, $\gamma>0$.
Since our focus is on threshold exceedances, we recall that $F \in \mathcal{D}(G_\gamma)$ also implies that, for any $t < x_E$,
\begin{equation}\label{eq:GP}
\lim_{t\to x_E}\Prob\left(\frac{X_i-t}{a(t')} \leq z \mid X_i>t\right)=H_\gamma(z):=
1-\left(1+\gamma z\right)_+^{-1/\gamma},
\end{equation}
where $H_\gamma$ is the GP distribution (e.g., \citealt[][Ch.1]{haan2006}, \citealt{balkema1974}), whose density is $h_{\gamma}(z)= \left(1+\gamma z\right)_+^{-1/\gamma-1}$, and  
$t'=1/(1-F(t))$. 

This result is useful in two key respects. First, for $x > t$, with $t$ close to $x_E$, the conditional distribution of $X_i$ given $X_i > t$ can be approximated by $H_{\boldsymbol{\vartheta}}(x - t)$, where $H_\bfvartheta(x)=H_\gamma(x/\sigma)$ with $\bfvartheta=(\gamma, \sigma)^\top$ and $\sigma$ is representative of $a(t')$. Its density is $h_\bfvartheta(x-t)$, where $h_{\bftheta}=h_\gamma(x/\sigma)/\sigma$. Second, for $\tau\in(0,1)$, let $Q (\tau) =  \inf( x : F(x) \leq \tau)$ be the generalised quantile function of $F$.  For $x>t$ the unconditional distribution $F(x)$ can be approximated as $F(x)\approx F(t) + \overline{F}(t)H_\bfvartheta(x)$, where $\overline{F}=1-F$. Combining this with $\tau=F(x)$, one obtains the approximation of tail quantiles, for $\tau\to1$ by
\begin{equation}\label{eq:ext_Q_pot}
Q(\tau)\approx t + \sigma\frac{(t'(1-\tau))^{-\gamma}-1}{\gamma}.
\end{equation}
\begin{rem}\label{cond:overall_conditions}
$H_\mathbf{\bfvartheta}$ is a useful model for inference when the true distribution $F$ is unknown. However, as it is statistically misspecified for any fixed $t$, regardless of how large $t$ is, such inference may suffer from bias. A standard approach in the extreme value literature to control such an estimation bias  is to introduce the Second-Order Condition (SOC), which essentially details that the convergence in \eqref{eq:GP} to $H_\gamma(z)$ occurs at rate $\mathcal{A}(t')$, where $\mathcal{A}$ is a regularly varying \textit{rate function} of index $\rho\leq 0$. The theoretical results of our proposed inference rely on the SOC, together with some assumptions on the temporal dependence structure, such as $\beta$-mixing and certain regularity conditions on the pairwise tail copulas. We collectively refer to these assumptions as the Temporal Dependence Conditions (TDC). Nonetheless, our methodology also remains applicable under the Unbiased First-Order Condition (UFOC; similar to that originally proposed by \citealp{drees2003}), which is considerably weaker than the SOC. We emphasize that these conditions are standard in the literature (see, e.g., \citealp[Ch. 2]{haan2006}; \citealp{drees2003}; \citealp{de2016adapting}; \citealp{chavez2018extreme}; \citealp{girard2021extreme}; \citealp{davison2023tail}) and are not restrictive, as they accommodate many popular time-series models. Due to space constraints, the detailed formulations of these conditions are provided in Section A of the Supplement (Conditions A.1–A.3). The theoretical results for the Bayesian inference presented in Section \ref{sec:Bayesian} are established under the Prior Distribution Conditions (PDC; see Condition A.4), which place only mild requirements on the families of prior distributions. In particular, they allow for widely used classes of priors, including informative, noninformative, and data-dependent specifications.
\end{rem}
%

%
\subsection{Preliminary results on likelihood-based inference}\label{sec:main}
%

Let $\bfX_n = (X_1, \ldots, X_n)$ be a sample from the stationary time series $(X_i)_{i \geq 1}$ with stationary cdf $F_0$, where $F_0 \in \mathcal{D}(G_{\gamma_0})$. The GP approximation for the distribution of threshold exceedances provides the theoretical foundation 
for the most widely used statistical method in the univariate extreme value setting \citep[e.g.][]{haan2006}. According to this approximation, exceedances $X_i$ from $\bfX_n$, conditionally on $X_i > t$ for some $i \in \{1, \ldots, n\}$, are approximately distributed as $H_{\boldsymbol{\vartheta}}(x - t)$.
We show that the following likelihood function and its associated inference, although originally designed for independent data, still retains desirable properties when applied in the time series context. Consider the independence pseudo loglikelihood for $\boldsymbol{\vartheta}$:
\begin{equation}\label{eq:logllik}
\mathscr{L}_n(\bfvartheta)=\frac{1}{k}\sum_{i=1}^k  \ell_{\bfvartheta}(X_{n-i+1,n}-X_{n-k,n}),
\end{equation}
where if $1 + \gamma x/\sigma > 0$, $\ell_{\boldsymbol{\vartheta}}(x) = -\log \sigma - \log h_\gamma(x/\sigma)$ is the GP log-likelihood for a single observation and $\ell_{\boldsymbol{\vartheta}}(x) =-\infty$, otherwise. Here, $X_{1,n} \leq \cdots \leq X_{n,n}$ denote the order statistics, and the top $k$ values $X_{n-k+1,n}, \ldots, X_{n,n}$ correspond to exceedances above the $(n-k)$th order statistic $X_{n-k,n}$, which is an estimator of the high threshold $Q_0(\tau_I)$. From a theoretical view point we assume $k=k_n$ with $k\to\infty$ as $n\to\infty$ and $k=o(n)$. Then, $Q_0(\tau_I)$ is considered an high ``in-sample'' threshold, provided that $\tau_I = \tau(n) \to 1$ as $n\to\infty$, and this is achieved by setting $\tau_I = 1-k/n$. Under this regime, $a_0(1/ \overline{F}(Q_0(\tau_I)))\approx a_0(1/ \overline{F}_n(X_{n-k,n}))=a_0(n/k)$, where $F_n$ is the empirical cdf. The Fisher information matrix associated with $\ell_{\bfvartheta}$, evaluated at the true parameter $\boldsymbol{\vartheta}_0$, is positive definite for $\gamma_0 > -1/2$. Hence, in the sequel we assume that $\bfvartheta_0\in(-1/2, \infty) \times (0, \infty)$ (see Supplement, Sections A and B).

The asymptotic properties of $\mathscr{L}_n(\bfvartheta)$ and of the corresponding maximum likelihood estimator (MLE) $\widehat{\bfvartheta}_n$ are more delicate to derive than in standard settings, since $\sigma$, which represents $a_0(n/k)$, is not fixed but depends on $n$. To address this issue, it is convenient to adopt the reparametrisation $\bftheta=\rho(\bfvartheta)=(\gamma, \sigma/a_0(n/k))^\top$, for $\bfvartheta\in\Theta$, and to focus on the reparametrised version of the log-likelihood $L_n(\bftheta)=k^{-1}\sum_{i=1}^k \ell_{\bftheta}((X_{n-i+1,n}-X_{n-k,n})/a_0(n/k))$, see \citet{dombry2023} for the independence case. In this case  $\boldsymbol{\theta}_0 = (\gamma_0, 1)^\top$. Once the asymptotic behavior $\widehat{\boldsymbol{\theta}}_n$  is established, that of $\widehat{\boldsymbol{\vartheta}}_n = \rho^{-1}(\widehat{\boldsymbol{\theta}}_n)$ follows as a byproduct. Our theory to be valid we need to work with the restricted parameter space $\Theta:=(-1, \infty) \times (0, \infty)$, since $L_n(\bftheta)$ is unbounded outside of it.

To present the key likelihood-based properties in the time series framework, we introduce some notation. For $\bftheta \in \Theta$, let $\bfI(\bftheta)$ denote the Fisher information associated with $\ell_{\boldsymbol{\theta}}$ (see Supplement, Section E) and $\bfI_0=\bfI(\bftheta_0)$. Define the rescaled score and negative observed information functions as $\bfS_n(\bftheta)=(\partial /\partial \bftheta) L_n(\bftheta)$ and $\bfJ_n(\bftheta)=(\partial^2/(\partial \bftheta \partial \bftheta^\top)) L_n(\bftheta)$.
For $\bftheta_0 \in \Theta$, let $\mathbb{B}_\epsilon(\bftheta_0) = \{\bftheta \in \Theta : \|\bftheta - \bftheta_0\| < \epsilon\}$ denote the ball of radius $\epsilon > 0$ centered at $\bftheta_0$, and $\mathbb{B}_\epsilon^\complement(\bftheta_0)$ its complement. Finally, we denote by $\mathcal{N}(\bfmu, \bfSigma)$ the multivariate normal distribution with mean vector $\bfmu$ and covariance matrix $\bfSigma$ (simplifying to $\mathcal{N}(\mu,\sigma^2)$ in the univariate case), and by $\mathcal{N}(\cdot; \boldsymbol{\mu}, \boldsymbol{\Sigma})$ the corresponding probability measure.
\begin{theo}\label{theo:llik_local_global_bounds}
Under the SOC and TDC, we have that $\bfS_n(\bftheta_0) = O_{\Prob}(1/\sqrt{k})$ and at the local level there is a $\epsilon,c_0>0$ such that with probability tending to $1$ as $n\to\infty$
\begin{equation*}\label{eq:Taylor_negative}
L_n(\bftheta) - L_n(\bftheta_0) \leq (\bftheta-\bftheta_0)^\top {\bfS_n(\bftheta_0)}-c_0 \|\bftheta-\bftheta_0\|^2,\quad \bftheta\in \mathbb{B}_{\epsilon}(\bftheta_0),
\end{equation*}
%
%
%
%
%
%
At the global level, there are $c_2,c_4>0$, $c_3  \in\Real$ and, for all large enough $\overline{\varrho}>1$, there is a $c_1>0$ such that, with probability tending to $1$ as $n\to\infty$ we have:
\begin{align*}
 L_n(\bftheta) - L_n(\bftheta_0) &\leq -\log\sigma-\frac{f(\varrho)}{\sigma}-v(\varrho), \quad \forall\,\bftheta\in\Theta,\\
L_n(\bftheta) - L_n(\bftheta_0) &\leq -c_4,\quad \mbox{if } \bftheta\in  \mathbb{B}_{\epsilon}(\bftheta_0)^\complement,
\end{align*}
with $\varrho=\gamma/\sigma$, $f(\varrho)=c_1\mathds{1}_{\varrho\leq \overline{\varrho}, \gamma> -1/2}+(2\varrho)^{-1}\log \varrho \mathds{1}_{\varrho> \overline{\varrho}}$, $v(\varrho)=-c_2\mathds{1}_{\varrho\leq \overline{\varrho}}+(\log \varrho+c_3)\mathds{1}_{\varrho> \overline{\varrho}}$.
The above results remain valid if \emph{SOC} is replaced by \emph{UFOC}(a).
\end{theo}
We point out that the log-likelihood local bounds are useful to determine the contraction rates
of the local MLE, computed over a narrow neighborhood of the true parameter as $\widehat{\bftheta}_n \in \argmax_{\mathbb{B}_{\epsilon}(\bftheta_0)} L_n(\bftheta)$, whose consistency and uniqueness follow as a byproduct. The likelihood global upper bounds are useful to establish that global MLE, computed over the entire parameter space  as $\widehat{\bfvartheta}_n \in \argmax_{\bfvartheta\in\Theta} \mathscr{L}_n(\bfvartheta)$, is actually the unique global likelihood maximiser.
\begin{cor}\label{cor:mle_uniqueness}
Under the conditions of Theorem \ref{theo:llik_local_global_bounds}, for $r\to\infty$ such that $r=o(\sqrt{k})$, we have $\|\widehat{\bftheta}_n-\bftheta_0\|=O_{\Prob}(r/\sqrt{k})$. Then, the sequence $(\widehat{\bftheta}_n)_{n\geq 1}$ of local MLEs  satisfies $\widehat{\gamma}_n=\gamma_0+O_{\Prob}(1/\sqrt{k})$ and $\widehat{\sigma}_n=a_0(n/k)\big(1+O_{\Prob}(1/\sqrt{k})\big)$ and is unique with probability tending to one, as $n\to\infty$.
The MLE $\widehat{\bfvartheta}_n$ is the unique global likelihood maximiser of $\mathscr{L}_n({\bfvartheta})$ over $\Theta$, with probability tending to one, as $n\to\infty$. 
\end{cor}
Global bounds also play a key role in deriving the asymptotic properties of the Bayesian procedure discussed in Section \ref{sec:Bayesian}. To complement this, the next result on the asymptotic normality of the local MLE clarifies how the serial dependence affects likelihood inference based on \eqref{eq:logllik}. For $i=1,2,\ldots$, the extremal dependence in the time series between different time points is measured by the tail copula function $R_i(x,y)$, defined for all $(x,y)\in[0,\infty]^2\setminus\{(\infty,\infty)\}$ as the limit of $t'\Prob(t' \overline{F}_0(X_1)\leq x, t' \overline{F}_0(X_{1+i})\leq y)$ as $t'\to\infty$. The quantity $R_0(x,y)=\min(x,y)+\sum_{i=1}^{\infty}(R_i(x,y)+R_i(y,x))$, i.e. the covariance function of a certain centered Gaussian
process (see \citealp[][]{drees2003} and Section D of the Supplement), is useful instead to represents  
the change of the MLE asymptotic variance from the independence case to
the mixing setting.
\begin{prop}\label{theo:loglike_expansion}
Under the conditions of Theorem \ref{theo:llik_local_global_bounds} we have
$\sqrt{k}(\widehat{\bftheta}_n-\bftheta_0)\stackrel{d}{\rightarrow} \mathcal{N}(\bfmu, \bfSigma_0)$, with $\bfmu$ given in Section E.3.2 of the Supplement, $\bfSigma_0=\bfSigma(\gamma_0,R_0)$ and
\begin{equation}\label{eq:asym_covariance}
\bfSigma(\gamma_0,R_0)=(1+\gamma_0)^2
\left(
\begin{matrix}
R_0(1,1) & \int_0^1\frac{R_0(u,1)}{u}\diff u -\frac{(2+\gamma_0)R_0(1,1)}{(1+\gamma_0)}\\
\int_0^1\frac{R_0(u,1)}{u}\diff u -\frac{(2+\gamma_0)R_0(1,1)}{(1+\gamma_0)} & 
\frac{(2+\gamma_0)^2R_0(1,1)}{(1+\gamma_0)^2}-2\frac{\int_0^1 \frac{R_0(u,1)}{u}\diff u}{1+\gamma_0}
\end{matrix}
\right).
\end{equation}
\end{prop}
Importantly, although $L_n(\boldsymbol{\theta})$ is an independence likelihood, the covariance matrix $\bfSigma_0$ still accounts for serial dependence. In the absence of serial dependence, the covariance function reduces to $R_0(x,y) = \min(x,y)$, so that $u^{-1} R_0(u,1) = 1$ and $R_0(1,1) = 1$. Consequently, $\bfSigma_0$ simplifies to the covariance structure obtained under independence (see \citealt[][Prop. 2.6]{dombry2023} and \citealt{d+f+d03}). Note that the covariance of $\sqrt{k}(\widehat{\bfvartheta}_n-\bfvartheta_0)$ is  approximately $\bfOmega_0=\bfOmega(\bfvartheta_0,R_0)=\bfA_0 \bfSigma(\gamma_0,R_0) \bfA_0^\top$ as $n\to\infty$, where $\bfA_0$ that is a diagonal matrix with elements $(1, a_0(n/k))$.

A crucial step in constructing confidence regions for $\boldsymbol{\vartheta}_0$ is the estimation of $\bfSigma_0$. Following \citet[][Proposition 2.1]{drees2003}, the covariance function $R_0(x,y)$ can be expressed as the limit of the suitably standardised covariance of the numbers of exceedances above different high quantiles of $F_0$. We therefore propose to estimate it using its empirical counterpart. Specifically,
$$
\widehat{R}_n(x,y)=\frac{n}{mk}\frac{1}{N}\sum_{j=1}^N \left(Z_j(x)-\frac{1}{N}\sum_{j=1}^NZ_j(x)\right)
\left(Z_j(y)-\frac{1}{N}\sum_{j=1}^NZ_j(y)\right),
$$
where  $N=n-m-1$ or $N=\lfloor n/(m+l) \rfloor$ and 
$$
Z_j(x)=\sum_{i=j}^{j+m-1}\indic\left(\overline{F}_n(X_i)\leq 1-x\frac{k}{n}\right),\, \text{ or }\, 
Z_j(x)=\sum_{i=1+(j-1)(m+l)}^{m+(j-1)(r+l)}\indic\left(\overline{F}_n(X_i)\leq 1-y\frac{k}{n}\right)
$$
in the sliding-block or disjoin-block approaches, respectively, and where $m=m_n$ and $l=l_n$ are indices satisfying $m\to\infty$ and $l\to\infty$ as $n\to\infty$ and $m=o(n/k)$ and $l=o(m)$. Then, we propose to estimate $\bfSigma_0$ by the estimator  $\widehat{\bfSigma}_n=\bfSigma(\widehat{\gamma}_n, \widehat{R}_n)$, where $\widehat{R}_n$ is defined above and
%
%
%
%
%
%
%
%
%
%
$\widehat{\gamma}_n$ is the MLE of $\gamma_0$. Accordingly, we propose to estimate $\bfOmega_0$ by the estimator $\widehat{\bfOmega}_n=\widehat{\bfA}_n \bfSigma(\widehat{\bftheta}_n, \widehat{R}_n)\widehat{\bfA}_n^\top$, where $\widehat{\bfA}_n$ is a diagonal matrix with elements $(1, \widehat{\sigma}_n)$ and $\widehat{\sigma}_n$ is the MLE of $a_0(n/k)$.  Estimators $\widehat{\bfSigma}_n$ and $\widehat{\bfOmega}_n$ are consistent and the $(1-\alpha)$-asymptotic confidence region for $\bfvartheta_0$, for any $\alpha\in(0,1)$, given by the random ellipsoid
\begin{equation}\label{eq:random_ellip_theta}
\widehat{\mathscr{E}}_n(\alpha)=\widehat{\bfvartheta}_n  + \widehat{\bfOmega}_n^{1/2} \mathbb{B}_{\varepsilon}(\bfzero),
\end{equation}
where $\varepsilon=(\chi^2_{2,1-\alpha}/k)^{1/2}$, $\bfzero=(0,0)^\top$ and $\chi^2_{2,1-\alpha}$ is the $(1-\alpha)$-quantile of the chi-square distribution with $2$ degrees of freedom, is asymptotically honest, i.e. its coverage probability achieves asymptotically the nominal level. The same is for the equi-tailed confidence intervals
\begin{equation}\label{eq:CI_param}
I_{n,1}(\alpha)=[\widehat{\gamma}_n\pm k^{-1/2}z_{1-\alpha/2}\widehat{\Omega}^{1/2}_{n,11}],\quad
I_{n,2}(\alpha)=[\widehat{\sigma}_n \pm k^{-1/2} z_{1-\alpha/2}\widehat{\Omega}^{1/2}_{n,22}],
\end{equation}
where $\widehat{\Omega}_{n,11}$ and $\widehat{\Omega}_{n,22}$ are the entries of the main diagonal of $\widehat{\bfOmega}_n$, and $z_\tau$ is the $\tau$-quantile of the standard normal distribution.
\begin{theo}\label{theo:consistency_var}
Under the conditions of Theorem \ref{theo:llik_local_global_bounds}, we have 
$$
\|\widehat{\bfSigma}_n-\bfSigma_0\|=o_\Prob(1), \quad \|\bfA_0\widehat{\bfA}_n^{-1}-\bfone\|=o_\Prob(1).
$$
where $\bfone$ is the identity matrix. If under SOC we have $\sqrt{k}\mathcal{A}_0(n/k)=o(1)$, then  $ \Prob(\bfvartheta_0 \in \mathscr{E}_n(\alpha))=1-\alpha+o(1)$, and $\Prob(\gamma_0\in I_{n,1}(\alpha))=\Prob(a_0(n/k)\in I_{n,2}(\alpha))=1-\alpha + o(1)$, where $\mathcal{A}_0$ is the rate function with index $\rho_0\leq0$. 
\end{theo}

In practice, risk assessment for events more severe than those observed to date can be carried out by estimating the $\tau$-quantiles of $F_0$ in the extrapolation regime. This is achieved by selecting an extreme level $\tau_E = \tau(n)$ such that $\tau_E \to 1$ and $n(1-\tau_E)=\to c \in [0,\infty)$ as $n \to \infty$. In the sequel we assume that such a condition is satisfied. The most interesting case is when $c \leq 1$, in which case the quantile $Q_0(\tau_E)$ is expected to lie in a neighborhood of, or above, the largest observations available. Defining the ratio $p = p_n = (1-\tau_E)/(1-\tau_I)$, we have $p \to 0$ as $n \to \infty$. The smaller $p$ is, the further into the tail of the data distribution the extrapolation extends. An approximation of $Q_0(\tau_E)$ is provided by \eqref{eq:ext_Q_pot}, and its MLE-based estimator is given by
\begin{equation}\label{eq:quantMLE}
\widehat{Q}_n(\tau_E)=X_{n-k,n}+\widehat{\sigma}_n \frac{p^{-\widehat{\gamma}_n}-1}{\widehat{\gamma}_n}.
\end{equation}
\begin{cor}\label{cor:mle_ext_quantile}
Work under SOC and TDC. Assume also that $\gamma_0<0$ when $\rho_0=0$. If $-\log p=o(\sqrt{k})$ we have
\begin{equation*}
\frac{ \sqrt{k} (\widehat{Q}_n(\tau_E) - Q_0(\tau_E))}{a_{0}(n/k) q_{\gamma_0} (1/p)} 
\overset{d}{\rightarrow} 
\mathcal{N} \left(\mu, \Sigma \right),
\end{equation*}
where $q_{\gamma_0}(x)=\int_1^{x} v^{\gamma_0-1}\log v \diff v$ and $\mu, \Sigma$ are given in Section E.3.4 of the Supplement. Furthermore, let
\begin{equation}\label{eq:CI_quantile}
I_{n,3}(\alpha)=[\widehat{Q}_n(\tau_E)\pm k^{-1/2}\widehat{\sigma}_n q_{\widehat{\gamma}_n}(1/p)z_{1-\alpha/2}\widehat{\Sigma}_n^{1/2}],
\end{equation}
where $\widehat{\Sigma}_n$ is a consistent estimator $\Sigma$ (such the one proposed in Section E.3.4. of the Supplement). Then, if $\sqrt{k}\mathcal{A
}_0(n/k)=o(1)$, we have $\Prob(Q_0(\tau_E)\in I_{n,3}(\alpha))=1-\alpha + o(1)$.
The above results remain valid if \emph{SOC} is replaced by \emph{UFOC}.
\end{cor}
%

%
%
\section{Bayesian inference}\label{sec:Bayesian}
%
%
%
%
\subsection{Marginal tail quantile-based risk estimation}\label{sec:static_inference}
%
%

We present a Bayesian procedure for risk assessment that relies on extrapolating tail quantiles of the marginal stationary distribution $F_0$ of the time series. Our starting point is inference on $\boldsymbol{\vartheta}$. 
We define a broader class of prior distributions by considering all prior joint densities $\pi(\boldsymbol{\vartheta})$ that satisfy the PDC (see Section A.2 of the Supplement). These conditions are relatively weak, and the resulting class includes many widely used priors: classical informative priors (see Examples A.5 in the Supplement), informative data-dependent priors, and standard non-informative priors \citep[see also the examples in][Sec. 2.3]{dombry2023}.

Then, the pseudo-posterior distribution is defined as
$$
\varPi_n(B)= \frac{\int_B \exp(k \mathscr{L}_n(\bfvartheta))\pi(\bfvartheta)\diff \bfvartheta}{\int_\Theta \exp(k \mathscr{L}_n(\bfvartheta))\pi(\bfvartheta)\diff \bfvartheta},
$$
for all measurable sets $B\subset \Theta$. We recall that, similar to frequentist inference of Section \ref{sec:main}, the asymptotic properties of the posterior distribution are derived by exploiting the reparametrisation $\boldsymbol{\theta} = \rho(\boldsymbol{\vartheta})$. In this case, we work with the reparametrised pseudo-posterior distribution $\overline{\varPi}_n := \varPi_n \circ \rho^{-1}$, which is constructed with  $L_n(\boldsymbol{\theta})$ and the reparametrised prior density  $\overline{\pi}(\bftheta):=\pi(\rho^{-1}(\bftheta))|(\partial^2/(\partial \bftheta \partial \bftheta^\top))\rho^{-1}|$.

Under the conditions of Theorem \ref{theo:llik_local_global_bounds} and PDC, we establish that $\overline{\varPi}_n$ contracts around the true parameter $\bftheta_0$ at an appropriate rate, and that the Bernstein–von Mises (BvM) theorem applies as $n \to \infty$. Consequently, $\overline{\varPi}_n$ is asymptotically normal with mean $\sqrt{k},\bfI^{-1}_0\bfS_n(\bftheta_0)$ and covariance $\bfI^{-1}_0$. For brevity, the results are deferred to Section F.1 of the Supplement. Since the analytical form of $\overline{\varPi}_n$ is not available, the BvM result is particularly useful for constructing credible regions for $\bftheta_0$.
From Proposition E.1 of the Supplement, $\text{Cov}(\sqrt{k}\bfS_n(\bftheta_0))\approx\bfV_0=\bfV(\bftheta_0, R_0)$, as $n\to\infty$, which implies that $\text{Cov}(\sqrt{k}\bfI^{-1}_0\bfS_n(\bftheta_0))\approx \bfI^{-1}_0\bfV_0\bfI^{-\top}_0$, as $n\to\infty$. This coincides with $\bfSigma_0$, the asymptotic covariance of $\sqrt{k}(\widehat{\bftheta}_n - \bftheta_0)$.
Because the asymptotic covariance of $\overline{\varPi}_n$ is $\bfI^{-1}_0$ rather than $\bfSigma_0$, the coverage probability of credible regions based on $\overline{\varPi}_n$ cannot, even in the absence of bias, achieve the nominal level asymptotically. This means that a posterior distribution constructed from an independence likelihood cannot fully capture the additional uncertainty induced by temporal dependence.
%
%
\subsubsection{Constructing posterior distribution with honest credible regions}\label{sec:adjusted_post}
%
%

To address this issue, we explain how to construct a new class of posterior distributions that produces honest credible regions. The construction is as follows. Define the pseudo log-likelihood 
\begin{equation*}\label{eq:ADJ_lik}
\mathscr{L}^{\star}_n(\bfvartheta^\star):=
\begin{cases}
\mathscr{L}_n(\widehat{\bfvartheta}_n + \widehat{\bfD}_n(\bfvartheta^\star_n-\widehat{\bfvartheta}_n)),\quad \bfvartheta^\star \in\varTheta_n,\\
-\infty, \hspace{9.2em} \text{otherwise},
\end{cases}
\end{equation*}
where $\widehat{\bfvartheta}_n$ is the MLE of $\bfvartheta_0$, $\varTheta_n:=\widehat{\bfvartheta}_n + \widehat{\bfD}^{-1}_n(\Theta-\widehat{\bfvartheta}_n)$, and 
where $\widehat{\bfD}_n=\widehat{\bfA}_n\widehat{\bfC}_n\widehat{\bfA}_n^{-1}$ 
with $\widehat{\bfC}_n$ that is an estimator of  $\bfC$ satisfying $\widehat{\bfC}_n=\bfC_0 + o_{\Prob}(1)$, and where $\bfC_0$ is a matrix  such that $\bfSigma_0=\bfC^{-\top}_0 \bfI^{-1}_0 \bfC^{-1}_0$. 
 Consider a generic joint prior density $\lambda(\bfvartheta^\star)$ defined on $\bfvartheta^\star\in \Real^2$. Prior distributions that can be used are: those induced by the map $\bfvartheta\mapsto \widehat{\bfvartheta}_n + \widehat{\bfD}^{-1}_n(\bfvartheta-\widehat{\bfvartheta}_n)$ and specific prior on $\bfvartheta^\star$ as informative or vague empirical Bayes ones, for concrete examples, see Examples A.6 and A.7 in the Supplement.

 Accordingly, we define, for all measurable sets $B\subset \Real^2$,  the new pseudo-posterior distribution
\begin{equation}\label{eq:ADJ_post}
\varPi_n^{\star}(B)= \frac{\int_B \exp(k \mathscr{L}^{\star}_n(\bfvartheta^\star))\lambda(\bfvartheta^\star)\diff \bfvartheta^\star}{\int_{\Real^2} \exp(k \mathscr{L}^{\star}_n(\bfvartheta^\star))\lambda(\bfvartheta^\star)\diff \bfvartheta^\star}.
\end{equation}

As usual for studying the theoretical behavior of $\varPi_n^{\star}$ we need to normalise $\bfvartheta^\star$ as we have done in the previous section, i.e. $\bftheta^\star=\rho(\bfvartheta^\star)=\bfA^{-1}_0\bfvartheta^\star$. 
The corresponding posterior is obtained applying the mapping $\overline{\varPi}^{\star}_n:=\varPi_n^{\star}\circ \rho^{-1}$. 
Natural questions are, what is the form of such a posterior distribution? And what are its properties? First, the corresponding prior density is $\overline{\lambda}(\bftheta^\star):=\lambda(\rho^{-1}(\bftheta^\star))|(\partial^2/(\partial \bftheta^\star \partial (\bftheta^\star)^\top))\rho^{-1}|$, which is a generic joint density on $\Real^2$. Second, the corresponding pseudo-log-likelihood is $L_n^{\star}(\bftheta^\star):=L_n(\widehat{\bftheta}_n + \bfC_n(\bftheta^\star-\widehat{\bftheta}_n))$, if $\bftheta^\star_n\in\Theta_n$, and $L_n^{\star}(\bftheta^\star_n)=-\infty$, if $\bftheta^\star_n\notin\Theta_n$, where $\bfC_n=\bfA^{-1}_0\widehat{\bfA}_n \widehat{\bfC}_n\widehat{\bfA}^{-1}_n\bfA_0$ and $\Theta_n:= \widehat{\bftheta}_n + \bfC^{-1}_n(\Theta-\widehat{\bftheta}_n)$. Then, the new family of pseudo-posterior distributions for the normalised parameter $\bftheta^\star$ is
\begin{equation}\label{eq:ADJ_post_rep}
\overline{\varPi}^{\star}_n(B)= \frac{\int_B \exp(k L_n^{\star}(\bftheta^\star))\overline{\lambda}(\bftheta^\star)\diff \bftheta^\star}
{\int_{\Real^2} \exp(k L_n^{\star}(\bftheta^\star))\overline{\lambda}(\bftheta^\star)\diff \bftheta^\star},
\end{equation}
for all measurable sets $B\subset \Real^2$. 

Note that the posterior in \eqref{eq:ADJ_post_rep} can be seen as a generalization of the posterior that is induced by the pseudo-posterior $\overline{\varPi}_n$ of the normalised parameter $\bftheta$, after applying the linear transformation 
$$
\bftheta \mapsto \bftheta^\star=\widehat{\bftheta}_n + \bfC^{-1}_n(\bftheta-\widehat{\bftheta}_n).
$$
This mapping ensures that credible regions based on the resulting posterior (i.e. $\overline{\varPi}_n^\star$) are asymptotically honest. Indeed, since $\bfC^{-1}_0=\bfI_C \bfSigma^\top_C$ then, we obtain
\begin{equation}\label{eq:decomposition}
\bfC^{-\top}_0 \bfI^{-1}_0 \bfC^{-1}_0 = \bfC^{-\top}_0 (\bfI_C \bfI^\top_C)^{-1} \bfC^{-1}_0=\bfSigma_C\bfI_C^{\top} (\bfI_C^{-\top} \bfI_C^{-1}) \bfI_C\bfSigma_C^\top=\bfSigma_C \bfSigma_C^\top=\bfSigma_0,
\end{equation}
where we recall that the Cholesky decomposition of a positive definite square matrix $\bfM$ is unique and is of the form $\bfM=\bfM_C \bfM_C^\top$, where $\bfM_C$ is a lower triangular matrix with real and positive diagonal entries. Therefore, given that $\bfC_n=\bfC_0+o_{\Prob}(1)$ the asymptotic variance of $\overline{\varPi}_n^\star$ is finally $\bfSigma_0$. 

We note that this type of transformation was previously proposed by \citet{chandler2007inference} to construct asymptotically honest confidence intervals based on the likelihood ratio statistic, when an independence likelihood is used in the time series context. Our work differs in several important respects. First, our new class of posteriors in \eqref{eq:ADJ_post_rep} includes the one obtained by the  linear mapping $\bftheta \mapsto \bftheta^\star$, only as a special case.
Second, we consider the more challenging setting of misspecified models, that remain so even in the absence of serial dependence. Third, the adjustment is required only in the Bayesian framework,
as no such adjustment is necessary in the frequentist setting, as confidence regions are already asymptotically honest, given the consistent estimation of $\bfSigma(\bftheta_0,R)$, established in Theorem \ref{theo:consistency_var}.
 \begin{theo}\label{theo:main_posterior}
Work under the SOC, TDC and PDC. Then, we have: 
\begin{enumerate}
\item \label{consistency} $\overline{\varPi}^{\star}_n$ is consistent with $\sqrt{k}$-contraction rate. There is a $M>0$ such that for all positive sequences $\epsilon=\epsilon_n\to0$, satisfying  $\sqrt{k}\epsilon \to \infty$ as $n\to\infty$,
$$
\overline{\varPi}^{\star}_n\left(B_{\epsilon}(\bftheta_0)^\complement\right)= O_\Prob\left(\exp\left(-Mk\epsilon^2\right)\right).
$$
\item \label{BvM} For all measurable sets $B\subset \Real^2$, as $n\to\infty$
$$
\sup_{B\subset \Real^2}|\overline{\varPi}^{\star}_n(\{\bftheta^\star:\sqrt{k}(\bftheta^\star-\bftheta_0)\in B\})-\mathcal{N}(B;\sqrt{k}\bfI^{-1}_0\bfS_{n}(\bftheta_0), \bfSigma_0|=o_\Prob(1).
$$
\item \label{coverage} On the basis of  point \ref{BvM}, we define for any $\alpha\in(0,1)$ the random ellipsoid 
\begin{equation}\label{eq:B_random_ellipsiod}
\widetilde{\mathscr{E}}_n(\alpha)=\widetilde{\bfmu}_n + \widetilde{\bfOmega}_n^{1/2} \bfB_2(\bfzero_2,(\chi^2_{2,1-\alpha})^{1/2}),
\end{equation}
where $\widetilde{\bfmu}_n$ and $\widetilde{\bfOmega}_n$ are the posterior mean and covariance of $\varPi_n^\star$, here seen as approximations of $\bfvartheta_0 + \bfA_0\bfI^{-1}_0\bfS_{n}(\bftheta_0)$ and $k^{-1}\bfOmega_0$. Furthermore, let
\begin{align}\label{eq:BCI_GPD}
\begin{split}
\widetilde{I}_{n,\gamma}(\alpha)&=[\varPi_{n,\gamma}^{\star\leftarrow}(\alpha/2),\, \varPi_{n,\gamma}^{\star\leftarrow}(1-\alpha/2)],\\
\widetilde{I}_{n,\sigma}(\alpha)&=[\varPi_{n,\sigma}^{\star\leftarrow}(\alpha/2),\, \varPi_{n,\sigma}^{\star\leftarrow}(1-\alpha/2)],
\end{split}
\end{align}
where $\varPi_{n,\gamma}^{\star\leftarrow}(p)$ and $\varPi_{n,\sigma}^{\star\leftarrow}(p)$ is the $p$-quantile of the univariate marginal posterior distributions of the individual parameters $\gamma$ and $\sigma$, respectively. If $\sqrt{k}\mathcal{A}_0(n/k)=o(1)$, then we have $ \Prob(\bfvartheta_0 \in \widetilde{\mathscr{E}}_n(\alpha))=1-\alpha+o(1)$, and $\Prob(\gamma_0\in \widetilde{I}_{n,\gamma}(\alpha))=\Prob(a_0(n/k)\in\widetilde{I}_{n,\sigma}(\alpha))=1-\alpha + o(1)$.
\end{enumerate}
The above results remain valid if \emph{SOC} is replaced by \emph{UFOC}(a).
\end{theo} 

Finally, let $\widetilde{\gamma}$, $\widetilde{\sigma}$ be distributed according to $\varPi^{\star}_n$ and consider the transformation $\widetilde{Q}_n(\tau_E)=X_{n-k,n}+\widetilde{\sigma}  \frac{p^{-\widetilde{\gamma}}-1}{\widetilde{\gamma}}$, which induces a posterior distribution, say $\Phi_n$ on the extreme quantile $Q(\tau_E)$. By ignoring the uncertainty associated with estimating the threshold via  $X_{n-k,n}$, the credible intervals derived from  $\Phi_n$ are not asymptotically honest.
To remedy this limitation we propose the refined transformation
\begin{equation}\label{eq:adjusted_quantile}
\mathscr{Q}_n(\tau_E)=
\widehat{Q}_n(\tau_E) + \widetilde{C}_n(\widehat{Q}_n(\tau_E)-\widehat{Q}_n(\tau_E)),
\end{equation}
where $\widetilde{C}_n^2=\widehat{\Sigma}_n/\widehat{V}_n$, and where $\widehat{\Sigma}_n$ and $\widehat{V}_n$ are consistent estimators of the asymptotic variances in Corollary \ref{cor:mle_ext_quantile} and of the posterior distribution $\Phi_n$, see Section F.2.2 of the Supplement for details. Transformation \eqref{eq:adjusted_quantile} induces a posterior distribution on the extreme quantile $Q(\tau_E)$, says $\varPhi_n^\star$, whose credible intervals are asymptotically honest.
\begin{cor}\label{cor:quantpost}
Work under SOC, TDC and PDC. Assume also that $\gamma_0<0$ when $\rho_0=0$. If $-\log p=o(\sqrt{k})$, then we have:
\begin{enumerate}
\item There is a $M>0$ such that for all positive sequences $\epsilon=\epsilon_n\to 0$, satisfying $\sqrt{k}\epsilon\to\infty$ and $-\epsilon\log(p)\to 0$ as $n\to\infty$,
$$
\varPhi^{\star}_n 
\left(\left\{ 
\mathscr{Q}_n(\tau_E):
\left|
\frac{\mathscr{Q}_n(\tau_E)-Q_0(\tau_E)}{q_{\gamma_0}(1/p)a_0(n/k)}\right|>\epsilon\right\} \right)= 
O_\bbP\left(e^{-Mk\epsilon^2}\right).
$$
\item For all measurable sets $B\subset \Real$ we have 
$$
\sup_{B\subseteq \Real}
\left|
\varPhi^{\star}_n
\left(
\sqrt{k}
\left(
\frac{\mathscr{Q}_n(\tau_E)-Q_0(\tau_E)}{q_{\gamma_0}(1/p)a_0(n/k_n)}
\right)\in B\right)-
\mathcal{N}(B; \Delta_n, \Sigma)
\right|
=o_\bbP(1),
$$
where $\Delta_n=\sqrt{k}(\widehat{Q}_n(\tau_E)-Q_0(\tau_E))( q(1/p)a_0(n/k))^{-1}$, with $\widehat{Q}_n(\tau_E)$ as in \eqref{eq:quantMLE}.
\item If $\sqrt{k}\mathcal{A}_0(n/k)\to 0$ as $n\to \infty$, we have for any $\alpha\in(0,1)$,
\begin{equation}\label{eq:QBCI_EQ}
\bbP\left(Q_0(\tau_E)\in \left[{\varPhi}_{n}^{\star\leftarrow}(\alpha/2); {\varPhi}_{n}^{\star\leftarrow}(1-\alpha/2)\right]\right)= 1-\alpha+o(1),
\end{equation}
where $\varPhi_{n}^{\star\leftarrow}(p)$ is the $p$-quantile of $\varPhi^{\star}_n$.
\end{enumerate}
The above results remain valid if \emph{SOC} is replaced by \emph{UFOC}.
\end{cor}
%

%
\subsection{Dynamic tail quantile-based risk estimation}\label{sec:Dyn_Bayesian}
%
%
We propose a simple yet effective approach for extrapolating tail events from time series in a dynamic framework by incorporating past information. To this end, we consider the class of heteroscedastic regression models $Y_i = v(\bfZ_i) + w(\bfZ_i)X_i$, where $v$ and $w>0$ are unknown measurable functions, $(\bfZ_i)$ may be partially or even completely unobserved, and $(X_i)$ is now a strictly stationary innovation sequence whose stationary distribution $F_0$ belongs to the domain of attraction $\DoA(G_{\gamma_0})$ with $\gamma_0 > -1/2$.
This family encompasses standard regression models (possibly misspecified due to temporal dependence in $(X_i)$), widely used time series models with independent innovations such as ARMA, ARCH, and GARCH, as well as models with dependent innovations, including ARMA–GARCH and continuous diffusion models with observations at discrete time such as Ornstein–Uhlenbeck (Example \ref{ex:HTAR}) and more general CARMA processes (Example \ref{ex:WARMA}(b)).

The key advantage of this model class is that, by homogeneity and equivariance of the quantile function, we obtain for any $\tau \in (0,1)$,
$$
Q_{Y_i\mid \bfZ_i}(\tau)=v(\bfZ_i) + w(\bfZ_i)Q_0(\tau).
$$

Let $s_n=o(n)$ be a sequence of nonnegative integers and set $\bar{n}=s_n+n$. Consider a sample $\bfY_{\bar{n}}$ and an out-of-sample random variable $Y_{\bar{n}+1}$ representing a future observation. If $v$ and $w$ were known and both the covariates $(\bfZ_i)$ and innovations $(X_i)$, $i = 1, \ldots, \bar{n}+1$, were fully observed, we could directly apply the extrapolation formula
$\widehat{Q}^{(\bar{n})}_{Y_{\bar{n}+1}|\bfZ_{\bar{n}+1}}(\tau_E)=v(\bfZ_{\bar{n}+1}) + w(\bfZ_{\bar{n}+1})\widehat{Q}_{\bar{n}}(\tau_E)$, where $\widehat{Q}_{\bar{n}}(\tau_E)$ denotes the estimator of the  innovations' extreme quantile $Q_0(\tau_E)$, obtained using the entire sample.
Innovations are however unobservable, and the covariates $\bfZ_i$ may be partly or entirely latent. We work then with the last $n$ residuals defined as
\begin{equation}\label{eq:residuals}
\widehat{X}_i^{(\bar{n})}=(Y_{i+s_n} -\widehat{v}_{i+s_n}^{(\bar{n})})/\widehat{w}_{i+s_n}^{(\bar{n})}, \quad i=1,\ldots,n,
\end{equation}
where $\widehat{v}_{i}^{(\bar{n})}$ and $\widehat{w}_{i}^{(\bar{n})}$ are suitably consistent estimators of $v(\bfZ_i)$ and $w(\bfZ_i)$ based on the entire sample $\bar{n}$. Then, we can show that Bayesian inference based on such residuals still satisfies desirable accuracy properties. We apply the Bayesian machinery described in Section \ref{sec:Bayesian} to the residuals in \eqref{eq:residuals}, deriving the corresponding adjusted posteriors \eqref{eq:ADJ_post_rep} and \eqref{eq:ADJ_post}. In the next theorem, the conditions and true GP parameters refer to the innovations $(X_i)$, while likelihood-based quantities (in particular, posterior distributions and credible regions) are constructed from the residuals $\widehat{X}_i^{(\bar{n})}$.
 \begin{theo}\label{theo:main_posterior_residuals}
Work under the SOC, TDC and PDC. Assume that residuals in \eqref{eq:residuals} satisfy
\begin{equation}\label{eq:residual}
\sqrt{k}\max_{0 \leq i \leq k}\frac{|\widehat{X}_{n-k+i,n}^{(\bar{n})}-X_{n-k+i,n}|}{{a_0(n/k)}}\stackrel{\Prob}{\to}0,\quad n\to\infty,
\end{equation}
where $\widehat{X}_{1,n}^{(\bar{n})} \leq \cdots \leq \widehat{X}_{n,n}^{(\bar{n})}$ and 
$X_{1,n}\leq \cdots \leq X_{n,n}$ are the order statistics of $(\widehat{X}_{i}^{(\bar{n})})$,  $i=1,\ldots,n$ and $(X_i)$, $i=s_n+1,\ldots,\bar{n}$, respectively.
Then, we have:
\begin{enumerate}
\item There is a $M>0$ such that for all  positive $\epsilon=\epsilon_n\to0$, satisfying  $\sqrt{k}\epsilon\to \infty$ as $n\to\infty$,
$$
\overline{\varPi}^{\star (\bar{n})}_n\left(B_{\epsilon}(\bftheta_0)^\complement\right)= O_\Prob\left(\exp\left(-Mk\epsilon^2\right)\right).
$$
\item For all measurable sets $B\subset \Theta$, as $n\to\infty$
$$
\sup_{B\subset \Theta}|\overline{\varPi}^{\star (\bar{n})}_n(\{\bftheta:\sqrt{k}(\bftheta-\bftheta_0)\in B\})-\mathcal{N}(B;\sqrt{k}\bfI^{-1}_0\bfS^{(\bar{n})}_{n}(\bftheta_0), \bfSigma_0)|=o_\Prob(1),
$$
\item For $\alpha\in(0,1)$, define the random ellipsoid $\widetilde{\mathscr{E}}_n(\alpha)$ and intervals $\widetilde{I}_{n,\gamma}(\alpha)$ and $\widetilde{I}_{n,\sigma}(\alpha)$ analogously to \eqref{eq:B_random_ellipsiod}--\eqref{eq:BCI_GPD}.
If $\sqrt{k}\mathcal{A}_0(n/k)\to 0$ as $n\to \infty$, then we have $ \Prob(\bfvartheta_0 \in \widetilde{\mathscr{E}}^{(\bar{n})}_n(\alpha))=1-\alpha+o(1)$, and $\Prob(\gamma_0\in \widetilde{I}_{n,\gamma}(\alpha))=\Prob(a_0(n/k)\in\widetilde{I}_{n,\sigma}(\alpha))=1-\alpha + o(1)$
\end{enumerate}
The above results remain valid if \emph{SOC} is replaced by \emph{UFOC}(a).
\end{theo} 
Let $\widetilde{\gamma}$, $\widetilde{\sigma}$ be the GP parameters distributed according the posterior $\overline{\varPi}^{\star (\bar{n})}_n$, applied to the residuals in \eqref{eq:residuals}. 
For the purposes of estimating the extreme quantile $Q_{Y_{\bar{n}+1}|\bfZ_{\bar{n}+1}}(\tau_E)$, concerning the out-of-sample random variable $Y_{\bar{n}+1}$, we consider the transformation
$$
\mathcal{Q}_{Y_{\bar{n}+1}\mid\bfZ_{\bar{n}+1}}(\tau_E)=\widehat{v}_{\bar{n}+1}^{(\bar{n})} + \widehat{w}_{\bar{n}+1}^{(\bar{n})}
\mathscr{Q}_n(\tau_E),
$$
where $\mathscr{Q}_n(\tau_E)$ is as in \eqref{eq:adjusted_quantile} but applied to the residuals,
which induces a posterior distribution on $Q_{Y_{\bar{n}+1}|\bfZ_{\bar{n}+1}}(\tau_E)$, that we denote by $\varPsi_n^{\star}$. Next result shows that accuracy properties of such a  posterior. We recall that the term $q_{\gamma_0}(\cdot)$ is next meant as in Corollary \ref{cor:mle_ext_quantile} and is derived from the quantile function of innovations' distribution $F_0$.
\begin{cor}\label{cor:quantpost_dynamic}
Work under Theorem \ref{theo:main_posterior_residuals} initial conditions and assume that $\gamma_0<0$ if $\rho_0=0$ and that
\begin{equation}\label{eq:dynquantcond}
\frac{|\widehat{v}_{\bar{n}+1}^{(\bar{n})}-v(\bfZ_{\bar{n}+1})|/\widehat{w}_{\bar{n}+1}^{(\bar{n})}}{a_0(n/k)q_{\gamma_0}(1/p) } = o_{\mathbb{P}}(1/\sqrt{k}), \quad \frac{|w(\bfZ_{\bar{n}+1})/\widehat{w}_{\bar{n}+1}^{(\bar{n})}-1| |Q_0(\tau_E)|}{a_0(n/k)q_{\gamma_0}(1/p) } =o_{\mathbb{P}}(1/\sqrt{k}),
\end{equation}
Then, if $-\log p = o(\sqrt{k})$, we have:
\begin{enumerate}
\item There is a $M>0$ such that for all $\epsilon=\epsilon_n\to 0$, satisfying $\sqrt{k}\epsilon\to\infty$ and $-\epsilon\log(p)\to 0$ as $n\to\infty$,
$$
\varPsi^{\star (\bar{n})}_n 
\left(\left\{ 
\mathcal{Q}_{Y_{\bar{n}+1}|\bfZ_{\bar{n}+1}}(\tau_E):
\left|
\frac{\mathcal{Q}_{X_{\bar{n}+1}|\bfZ_{\bar{n}+1}}(\tau_E)-Q_{Y_{\bar{n}+1}|\bfZ_{\bar{n}+1}}(\tau_E)}{q_{\gamma_0}(1/p)a_0(n/k)\widehat{w}_{\bar{n}+1}^{(\bar{n})}}\right|>\epsilon_n\right\} \right)= 
O_\bbP\left(e^{-Mk\epsilon^2}\right).
$$
\item For all measurable sets $B\subset \Real$ we have 
$$
\sup_{B\subseteq \Real}
\left|
\varPsi^{\star (\bar{n})}_n
\left(
\sqrt{k}
\left(
\frac{\mathcal{Q}_{Y_{\bar{n}+1}|\bfZ_{\bar{n}+1}}(\tau_E)-Q_{Y_{\bar{n}+1}|\bfZ_{\bar{n}+1}}(\tau_E)}{q_{\gamma_0}(1/p)a_0(n/k)\widehat{w}_{\bar{n}+1}^{(\bar{n})}}
\right)\in B\right)-
\mathcal{N}(B; \Delta_n, \Sigma)
\right|
=o_\bbP(1),
$$
where $\Delta_n$ is given in Section F.2.4 of the Supplement.
\item If $\sqrt{k}\mathcal{A}(n/k)=o(1)$, then for any $\alpha\in(0,1)$ we have
\begin{equation}\label{eq:QBCI_EQ_dynam}
\bbP\left(Q_{Y_{\bar{n}+1}|\bfZ_{\bar{n}+1}}(\tau_E)\in \left[{\varPsi}_{\bar{n}}^{\star (\bar{n})\leftarrow}(\alpha/2); {\varPsi}_{n}^{\star (\bar{n})\leftarrow}(1-\alpha/2)\right]\right)= 1-\alpha+o(1),
\end{equation}
where $\varPsi_{n}^{\star(\bar{n})\leftarrow}(\tau)$ is the $\tau$-quantile of $\varPsi^{\star (\bar{n})}_n$.
\end{enumerate}
The above results remain valid if \emph{SOC} is replaced by \emph{UFOC}(b).
\end{cor}
We conclude this section by presenting practical examples of time series that encompass a wide range of tail behaviors and dependence structures, for which our dynamic Bayesian inference satisfies the asymptotic results outlined above.
In particular, we refer to Corollary E.23 of the Supplement, which provides an example of the residual-based estimators $\widehat{\bfA}_n^{(\bar{n})}$, $\widehat{\bfSigma}_n^{(\bar{n})}$ and $\widehat{\bfOmega}_n$. When applied in the next examples, these estimators are consistent with our theoretical framework, provided that $s_n/\log n \to \infty$ as $n \to \infty$.
\begin{ex}\label{ex:HTAR}
The standard class of autoregressive processes with heavy-tailed iid innovations $(X_i)$, is obtained with 
$$
v(\boldsymbol{Z_i})= \sum_{j=1}^p \phi_j Y_{i-j}, \quad w(\boldsymbol{Z}_i) \equiv 1.
$$
Residuals can be constructed using \eqref{eq:residuals} for $i=1, \ldots, n$, with $\widehat{v}_{i+s_n}^{(\bar{n})} = \sum_{j=1}^p \widehat{\phi}_{\bar{n},j} Y_{i+s_n-j}$ and $\widehat{w}_{i+s_n}^{(\bar{n})} =1$. In this way, as soon as the estimators $\widehat{\phi}_{\bar{n},j}, \, 1 \leq j \leq p,$ are consistent with rate at least $n^{-\delta}$ for some $\delta>0$ (e.g., \citealp{D+K+L92}, Th. 2.3; \citealp{A+C+D2009}, Th. 3.2), then \eqref{eq:residual}--\eqref{eq:dynquantcond} are satisfied under SOC or UFOC, if $\gamma_0 \in (0,1)$ and the right-tail of $F_0$ is not heavier than the left one, $1-\sum_{j=1}^p \phi_j \neq 0$, for $|z|\leq 1$, and $k^{1/2+\gamma_0} =o(1){n}^\delta$. An important member of this class is the Ornstein-Uhlenbeck continuous diffusion processes specified by the equation $D Y(t)= - \zeta Y(t) + \varsigma D S(t) $, where $\zeta, \varsigma >0$  and $(S(t), \, t \geq 0)$ is a standard symmetric $\alpha$-stable L\'evy process, with $\alpha \in (0,2)$. Its discrete time version $Y_i=Y(i)$ has a AR$(1)$ representation where $\phi_1= e^{-\zeta}$ and $X_i= \varsigma \int_{i-1}^i e^{-\zeta(i-u)}\diff S(u)$ are iid variables with scaled symmetric $\alpha$-stable distribution.
\end{ex}
\begin{ex}\label{ex:WARMA}
The linear class of weak ARMA$(p,q)$ time series models is obtained with
 \begin{equation}\label{eq:ARMA}
 v(\bfZ_{i})= \sum_{j=1}^p \phi_j Y_{i-j}+ \sum_{j=1}^q \psi_j X_{i-j} , \quad w(\bfZ_i)\equiv 1, 
 \end{equation}
 where $\phi_j$ and $\psi_j$ are suitable coefficients and $(X_i)$ may be dependent over time \citep[][Sec. 2]{FRANCQ1998145}.  Popular members that exhibit stationary geometric beta mixing innovations are ARMA-(G)ARCH and CARMA (described below). 
 Starting with auxiliary residuals $\widehat{X}_{-s_n + 1}^{(\bar{n})}, \ldots,\widehat{X}_{0}^{(\bar{n})}$, where 
 $$
 \widehat{X}_{s_n+1}^{(\bar{n})}=\cdots=\widehat{X}_{-s_n+\max(p,q)}^{(\bar{n})}=0
 $$ 
 the residuals $\widehat{X}_i^{(\bar{n})}$, $i= 1, \ldots,n$, are iteratively obtained as in
 \eqref{eq:residuals}, where $\widehat{v}_{i + s_n}^{(\bar{n})}$, $i = -s_n + 1, \ldots, n$, itself is obtained by \eqref{eq:ARMA} with $\bfxi =(\phi_j, j=1,\ldots,p; \, \psi_j, j=1\ldots, q)$ replaced by a suitable estimator $\widehat{\bfxi}_{\bar{n}}$ and $X_{i + s_n - j}$ replaced by $\widehat{X}_{i - j}^{(\bar{n})}$, $j = 1, \ldots, q$, and with $\widehat{w}_{i}^{(\bar{n})}=1$.
 In addition to SOC (or UFOC), TDC and PDC,  sufficient conditions for our theory to apply are: $1+\sum_{j=1}^q \psi_j z^j \neq 0$ for $|z| \leq 1$, $\widehat{\bfxi}_{\bar{n}}$ is ${n}^{-\delta}$-consistent for some $\delta>0$ (e.g., \citealp{FRANCQ1998145}), there exists a solution 
 $$
 Y_i = \sum_{j =0}^\infty \rho_j X_{i-j}, \quad \sum_{j=0}^\infty|\rho_j| <\infty,\quad \sum_{j=0}^\infty |\rho_j X_{j}| =O_{\mathbb{P}}(1).
$$
and $\sqrt{k}(\max_{1l\leq i \leq \bar{n}}|X_i|+1)=o_\Prob(a_0(n/k)n^\delta)$ Two important members are the following.
\begin{itemize} 
\item The ARMA-(G)ARCH family is obtained when $(X_i )$ are of the form
 \begin{align*}
 X_i = \sqrt{h_i \eta_i}, \quad h_i= \beta_0  +\sum_{j=1}^r \beta_j X_{i-j}^2 + \sum_{j=1}^u \zeta_j  h_{i-j},
 \end{align*}
where $(\eta_i)$ are iid random variables such that $\mathbb{E}( \eta_i^2)=1$, $\beta_j>0$ for  $j=0,1,\ldots,r$, and $\zeta_j\geq0$ for $j=1,\ldots,u$ in the ARCH and GARCH case, respectively. 
The existence of a solution $Y_i$ and an example of  ${n}^{-\delta}$-consistent estimator $\widehat{\bfxi}_{\bar{n}}$ are discussed, e.g. in  \cite{Zhang2022},  
\item The L\'evy-driven CARMA$(p,q)$, $0 \leq q<p$, is the continuous time process defined as solution of the stochastic differential equation \citep[e.g.][]{Brockwell2014}
 $$
 Y(t)=- D^p Y(t) - \sum_{j=1}^{p-1} \alpha_j D^{p-j} Y(t)+ \beta_0 L(t) + \sum_{j=1}^{q-1} \beta_j D^{j} L(t) + D^q L(t)
 $$
where $D^j$ denotes $j$-th order differentiation over time, $\boldsymbol{\chi}=(\alpha_j, j=1,\ldots,p; \, \beta_j, j=0,\ldots,q-1)$ are real-valued coefficients and  $L$ is a L\'evy process. 
When $\mathbb{E}[L(1)] = 0$, $\mathbb{E}[L^2(1)] < \infty$, and suitable regularity conditions on $\boldsymbol{\chi}$ are satisfied \citep[][Section 7]{Brockwell2014}, the discretised process $Y_i = Y(i)$ admits a weak ARMA$(p,p-1)$ representation, where $(X_i)$ is an uncorrelated white noise sequence and the ARMA coefficients $\boldsymbol{\xi} \equiv \boldsymbol{\xi}(\boldsymbol{\chi})$ are linked to $\boldsymbol{\chi}$ as described in \citet[][Th. 2.1]{Brockwell2019}.
Under a sufficiently regular parametric specification $\boldsymbol{\chi}(\boldsymbol{\gamma})$ (e.g., \citealp{Schlemm2012EJS}, Cond. C.1–C.9), a preliminary consistent estimator $\widehat{\boldsymbol{\chi}}_{\bar{n}} = \boldsymbol{\chi}(\widehat{\boldsymbol{\gamma}}_{\bar{n}})$ (e.g., \citealp{Schlemm2012EJS}, Th. 3.16) leads to accurate estimation of the ARMA parameters via $\widehat{\boldsymbol{\xi}}_{\bar{n}} = \boldsymbol{\xi}(\widehat{\boldsymbol{\chi}}_{\bar{n}})$. A simpler alternative is to directly estimate the ARMA parameters \citep[][Sect. 3.2]{Garcia2011}. Although this approach does not provide an intrinsic finite-sample estimation of the ARMA representation of a CARMA process (since not every ARMA can be embedded in a CARMA), it asymptotically does so if the innovations satisfy suitable moment and mixing conditions \citep[][Th. 4.3]{Schlemm2012}.
This strategy yields a flexible dynamic framework for inference on tail quantiles. Since the focus is on tail behavior, modelling simplifications are less consequential than in other two-step procedures.
\end{itemize}
\end{ex}
\begin{ex}
Another important class is the ARMAX, obtained with 
$$
v(\bfZ_{i})= \sum_{j=1}^p \phi_i Y_{i-j}
+\sum_{j=0}^r \bfbeta_{j}^\top \bfE_{i-j}
+ \sum_{j=1}^q \psi_j X_{i-j}, \quad w(\bfZ_i)\equiv 1, 
$$
where $\bfE_i$ a random vector of observable exogenous explanatory variables. Residuals can be constructed as in the previous example by also  including  the estimated regression term $\sum_{j=0}^r \widehat{\bfbeta}_{\bar{n},j}^\top \bfE_{\bar{n}-j} $ in the linear term $\widehat{m}_{i}^{(\bar{n})}$.  Upon ${n}^\delta$-consistent estimation of  $\boldsymbol{\xi}$ and $\bfbeta_j$, $j=0, \ldots,r$ (e.g. in \citealp{Hannan1980}), our theory applies provided that $Y_i- \sum_{j=0}^r \bfbeta_j \bfE_{i-j}$ and $X_i$ 
satisfy the causal representation and bounds in the previous example, and $\max_{1 \leq i \leq  \bar{n}} \Vert \bfE_{i}\Vert = o_{\mathbb{P}}(a_0(n/k) n^\delta k^{-1/2})$.
\end{ex}
%

%
\section{Simulation study}\label{sec:simulations}
%
%
We implemented numerical experiments to evaluate the finite-sample performance of our Bayesian procedure proposed in Section \ref{sec:Bayesian}. In particular, we considered the symmetric Bayesian Adjusted Credible Region (BACR) estimator defined in \eqref{eq:B_random_ellipsiod}, as well as the Bayesian Adjusted Credible Interval (BACI) estimators in \eqref{eq:BCI_GPD} and \eqref{eq:QBCI_EQ}. For comparison, we included their unadjusted counterparts (BCR and BCI, respectively) and the likelihood-based estimators introduced in Section \ref{sec:main}, namely the symmetric Frequentist Confidence Region (FCR) estimator in \eqref{eq:random_ellip_theta} and the equi-tailed Frequentist Confidence Interval (FCI) estimators in \eqref{eq:CI_param} and \eqref{eq:CI_quantile}. The frequentist estimators require a preliminary estimate of the asymptotic covariance matrix $\bfSigma_0$, which we preliminary estimate by $\widehat{\bfSigma}_n$ (see the discussion below Proposition \ref{theo:loglike_expansion}). The performance of this estimator was also assessed in our experiments.
\begin{figure}[t!]
	\centering
	\includegraphics[page=4,width=.24\textwidth]{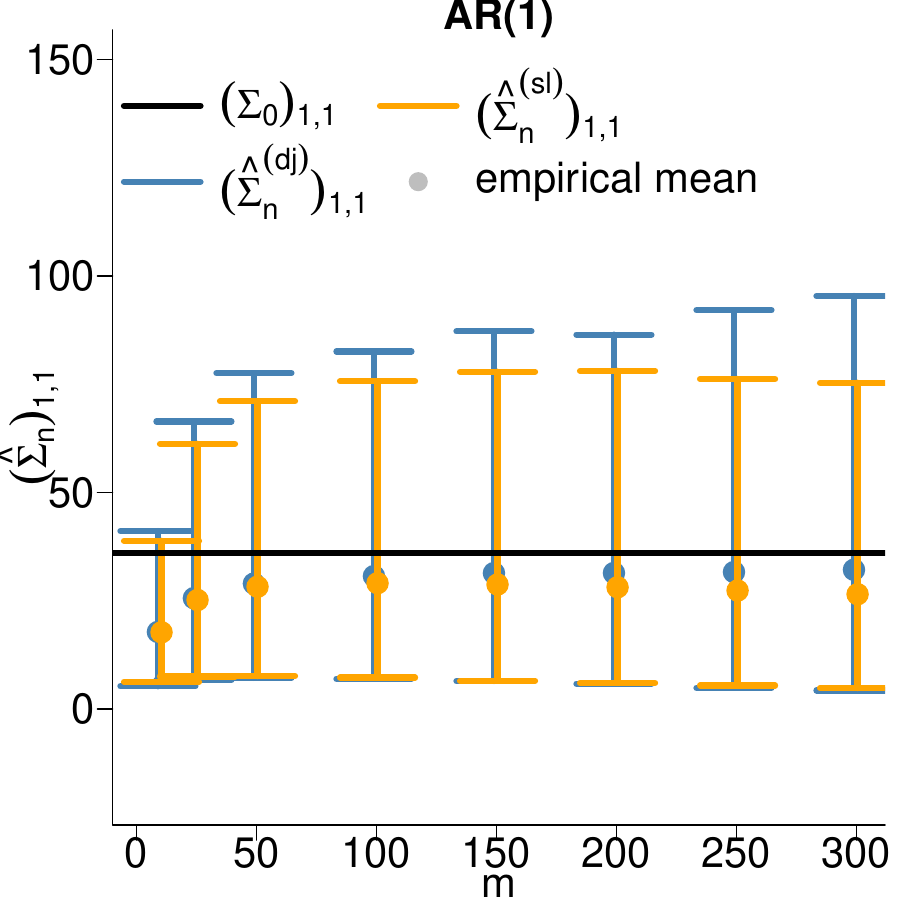}
	\includegraphics[page=6,width=.24\textwidth]{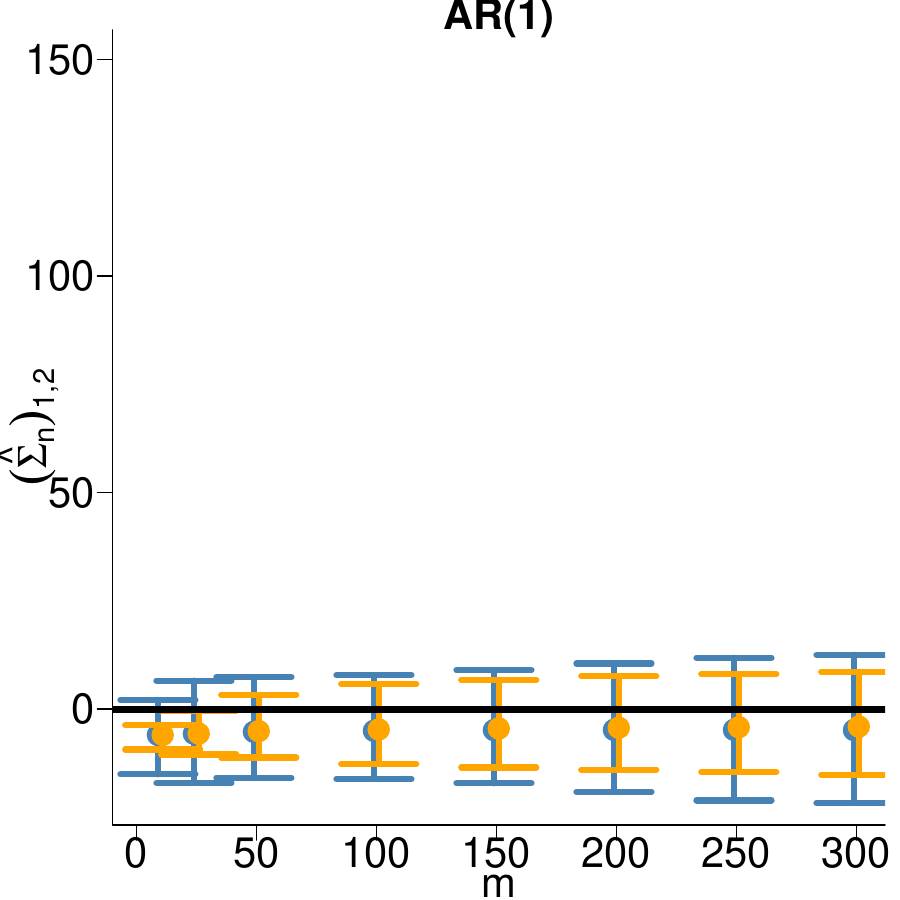}
	\includegraphics[page=5,width=.24\textwidth]{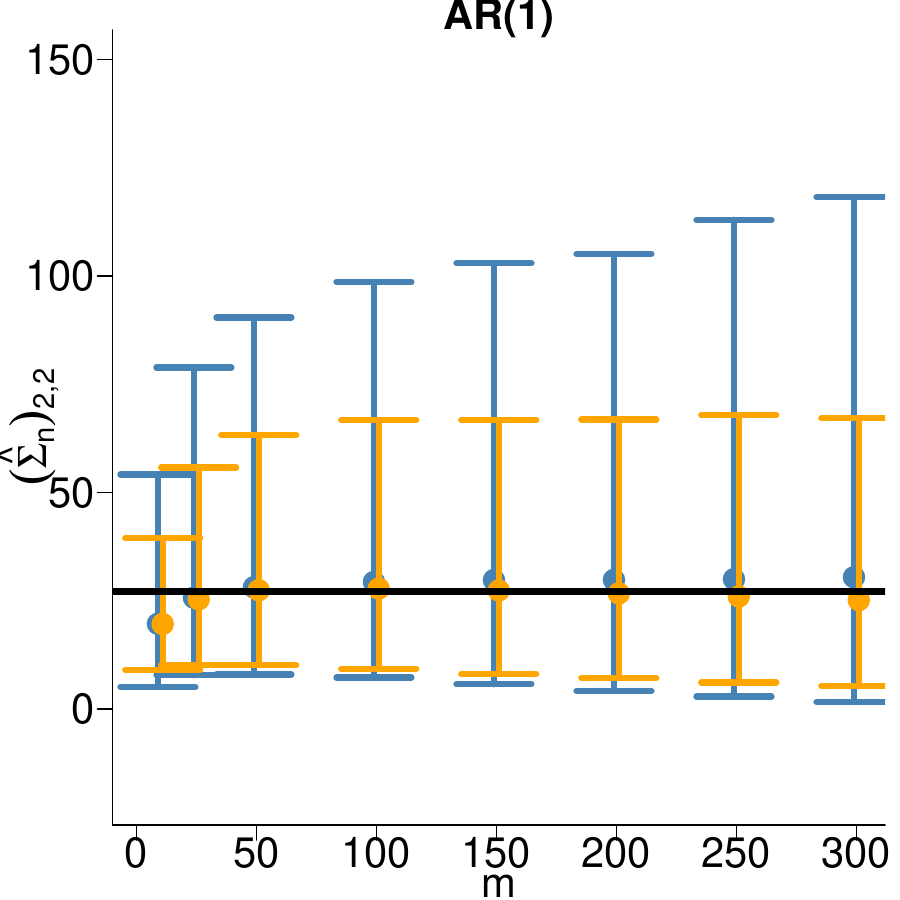}\\
	\includegraphics[page=7,width=.24\textwidth]{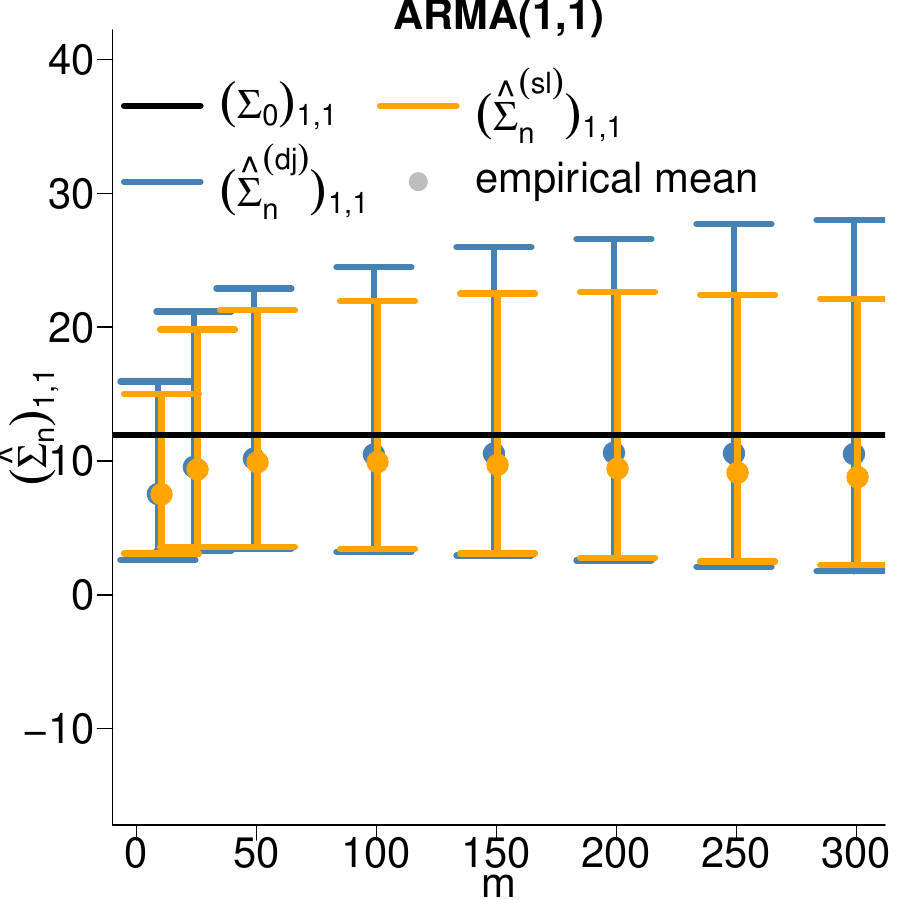}
	\includegraphics[page=9,width=.24\textwidth]{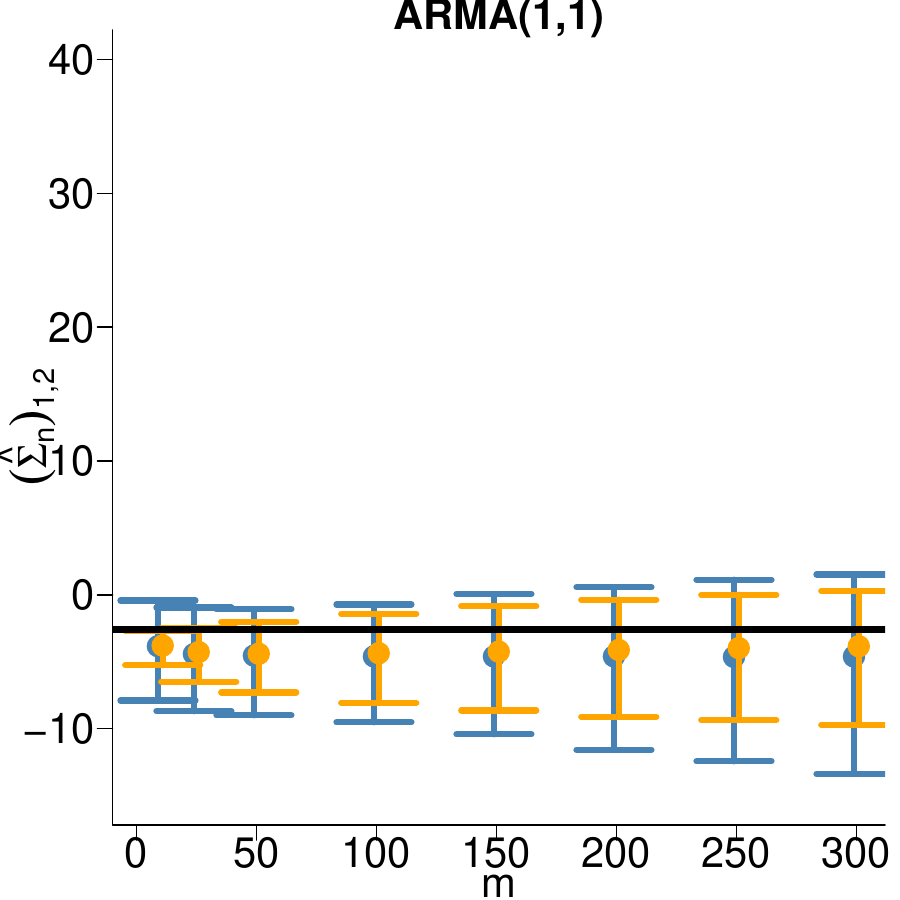}
	\includegraphics[page=8,width=.24\textwidth]{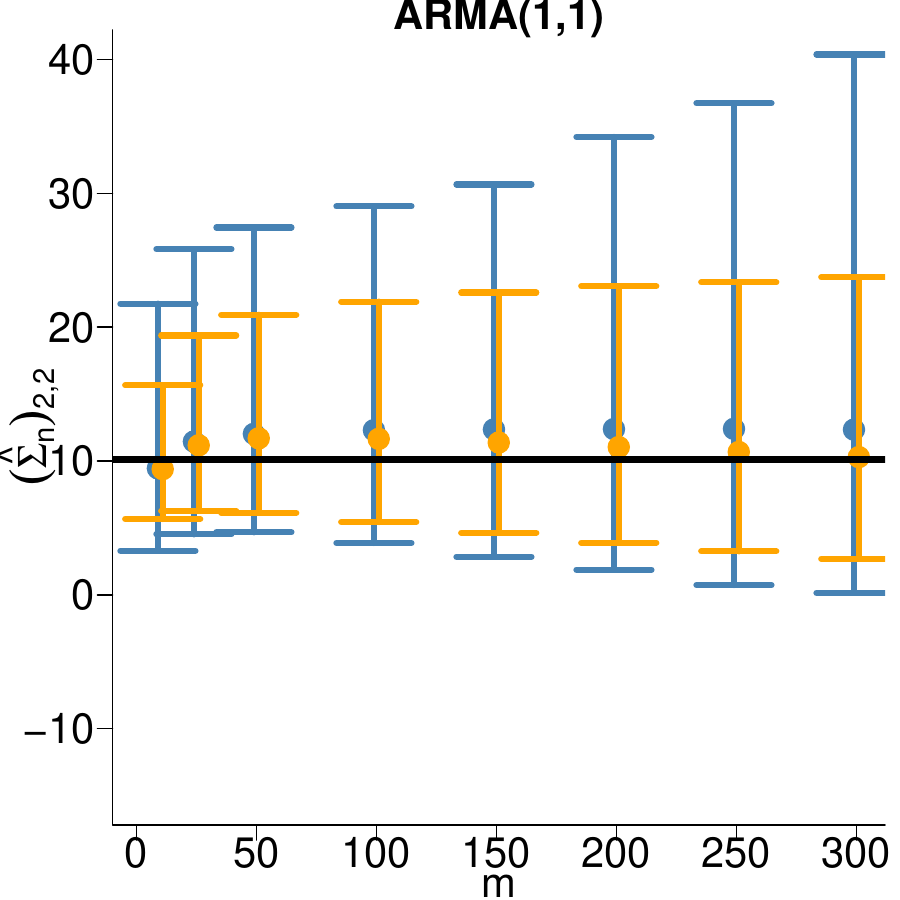}\\
	\includegraphics[page=10,width=.24\textwidth]{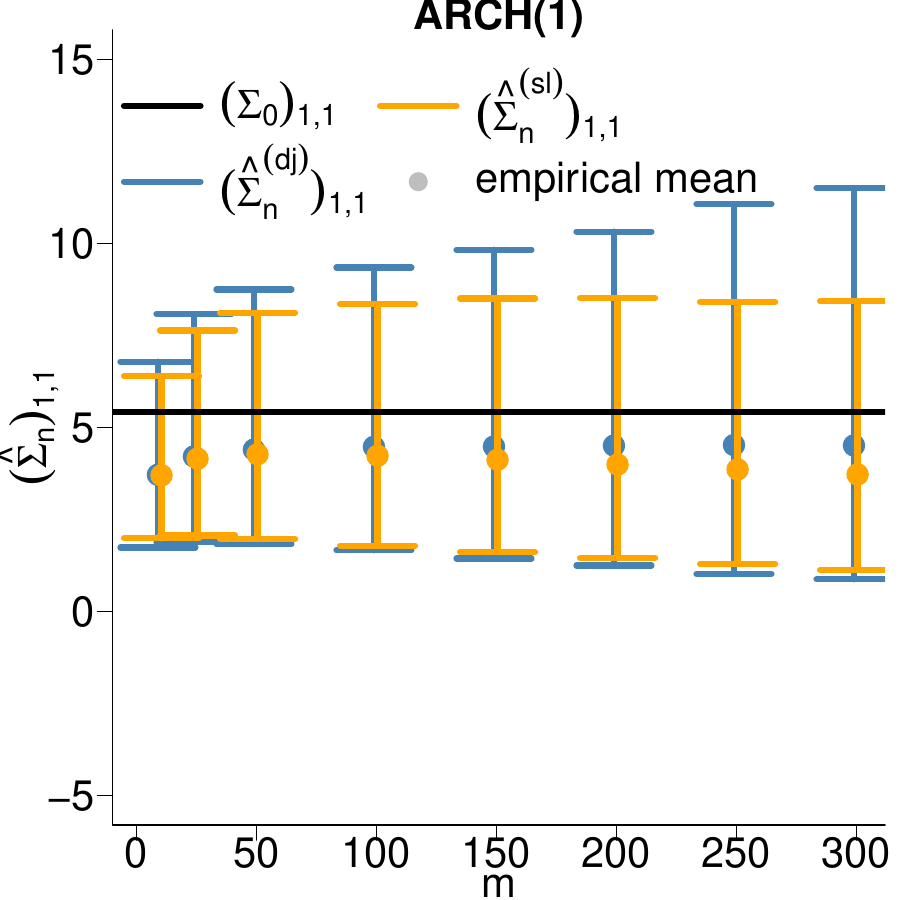}
	\includegraphics[page=12,width=.24\textwidth]{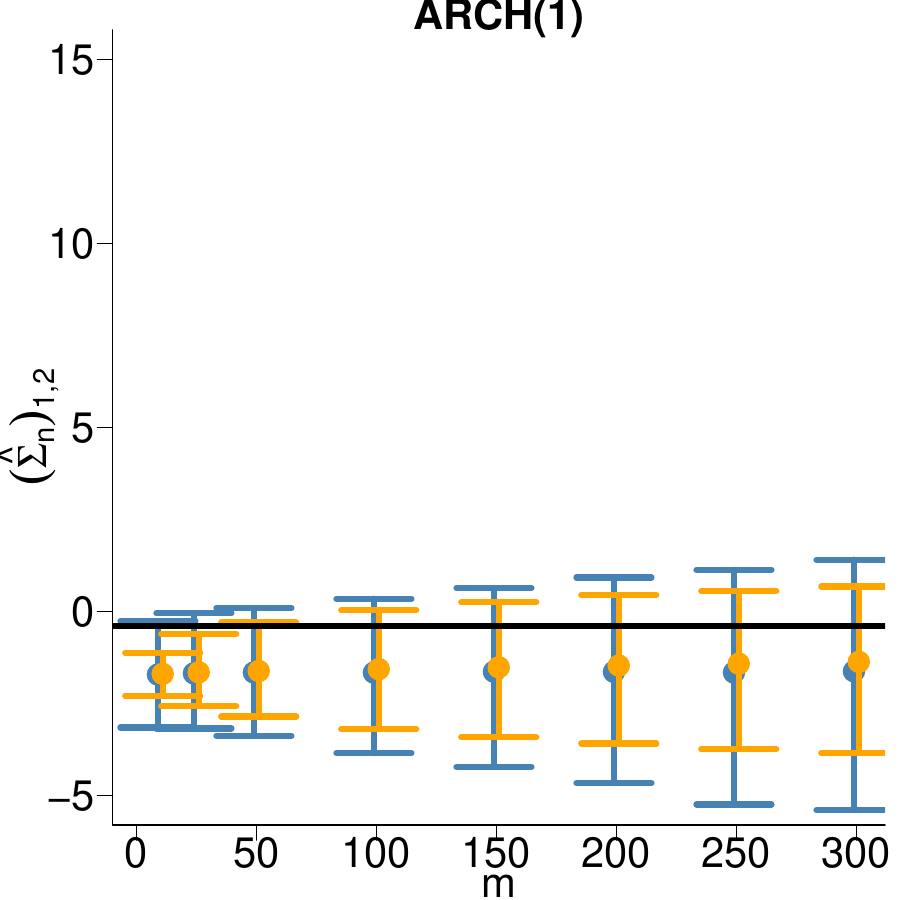}
	\includegraphics[page=11,width=.24\textwidth]{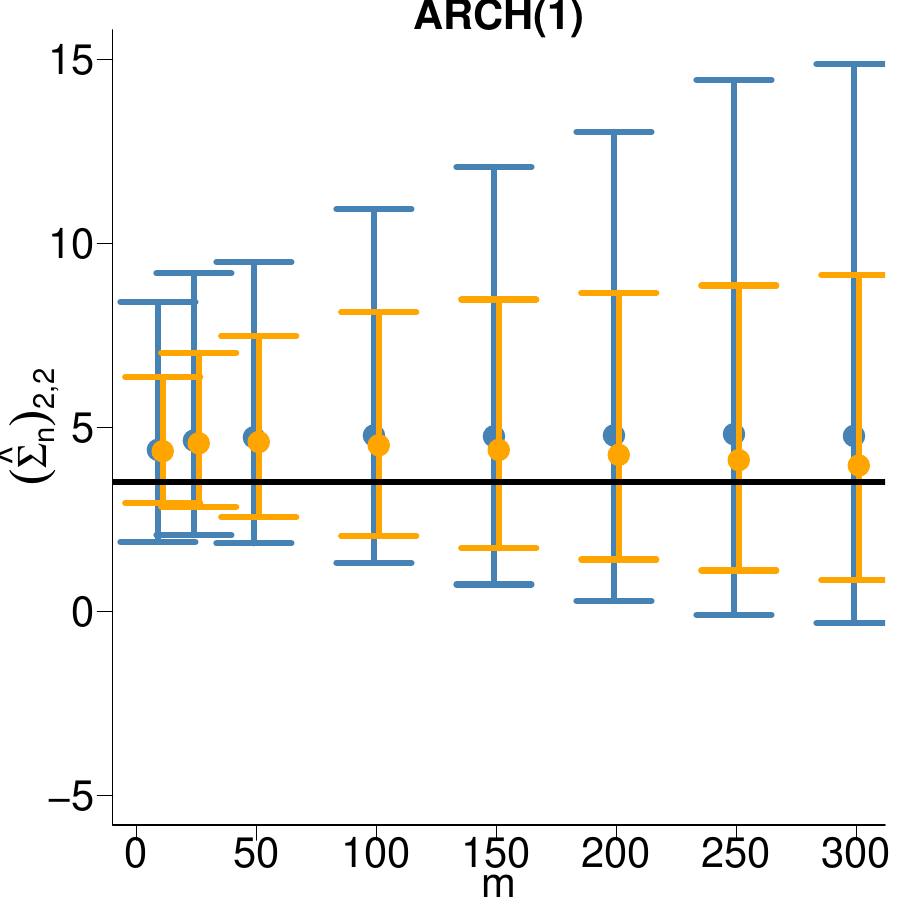}
	\caption{Mean values (dots) and 5\% and 95\% quantiles (horizontal lines) of the empirical distribution of the estimated diagonal components $(\bfSigma_0)_{11}$ (left) and $(\bfSigma_0)_{22}$ (right), and off diagonal element $(\bfSigma_0)_{12}$ (center) of the asymptotic covariance $\bfSigma_0$, computed across $N=5000$ simulations.}
\label{fig:sigma_estimator}
\end{figure}
Simulations are made using the following time series models:
\begin{enumerate}
\item[(a)] the AR(1) model $X_{i+1}=0.8X_i + \varepsilon_{i+1}$, where the innovations $\varepsilon_i$ are independent Student-$t$ distributed with $1$ degrees of freedom;
\item[(b)] the ARMA(1,1) model $X_{i+1}=0.8X_i + \varepsilon_{i+1} + 0.8\varepsilon_{i}$, where the innovations $\varepsilon_i$ are independent Student-$t$ distributed with $2$ degrees of freedom;
\item[(c)] the ARCH(1) model $X_{i+1}=\psi_{i+1}  \varepsilon_{i+1}$, where $\varepsilon_i$ are independent standard Gaussian innovations and $\psi_{i+1}^2 = 2\times 10^{-5} + 0.99 X_i^2$;
\item[(d)]  the Markovian Copula models $X_{i+1} = F_0^{-1} (U_{t+1})$, with $(1 - U_{i}, 1 - U_{i+1})$ following the Clayton Copula  $C_{\eta} (u_1, u_2) = (u_{1}^{-\eta} + u_{2}^{-\eta} - 1)^{-1 / \eta}$, where $\eta =(0.41, 1.06)$, and $F_0(x)=1-e^{-x}$ for $x\geq0$ in one case and $F_0(x)=1-(1-x)^3/9$ in the other,
\end{enumerate}
for the estimation of the GP parameters and tail quantiles pertaining to the stationary marginal distribution, and 
\begin{enumerate}
\item[(e)] the ARMA(2,1) model $Y_{i+1}=0.5 Y_i + 0.1875 Y_{i - 1} + X_{i+1} + 0.8X_{i}$, where the innovations $X_i$ are independent Student-$t$ distributed with $5$ degrees of freedom;
\item[(f)] the AR(1)-GARCH(1,1) model $Y_{i+1}=0.8Y_i + X_{i+1}$, where the innovations constitute a GARCH(1,1) model $X_{i+1} = \psi_{i+1}  \xi_{i+1}$ with $\psi_{i+1}^2 = 2\times 10^{-5} + 0.4 X_i^2 + 0.3 \psi_{i}^2$ and $\xi_i$ are independent standard Gaussian random variables,
\end{enumerate}
for the estimation of the GP parameters and dynamic, conditional on the past, tail quantiles.

The first, second, and fifth models are standard linear time series with strong serial dependence, and the extreme value indices of their marginal and innovation distributions are $\gamma_0 = 1$, $\gamma_0 = 1/2$, and $\gamma_0 = 1/5$, respectively. The marginal distribution of the third model and the innovation distribution of the sixth model are heavy-tailed, with extreme value indices $\gamma_0 \approx 0.493$ and $\gamma_0 \approx 0.207$, respectively, calculated using \citet{haan1989} and \citet[][Theorem 2.1]{mikosch2000limit}. The marginal distributions of the fourth model consist of (i) the unit-exponential distribution and (ii) a member of the power-law family $F(x) = 1 - K(x_E - x)^\nu$ with endpoint $x_E$, shape parameter $\nu > 0$, and constant $K > 0$ \citep[][Example 3.3.16]{embrechts1997modelling}. These belong to the domains of attraction of the GEV distribution with $\gamma_0 = 0$ and $\gamma_0 = -1/3$, respectively. Additional simulation results, including the independence case and models violating the conditions required for our results, are reported in Section B of the Supplement.

In the first experiment, we simulate $N = 5000$ independent time series from models (a)–(c), each of length $n = 2000$. For each replication, we estimate the asymptotic covariance matrix $\bfSigma_0$ using the estimator $\widehat{\bfSigma}_n$, computed with $k = 100$ and values of the index $m$ ranging from $10$ to $300$; see Proposition \ref{theo:loglike_expansion} and the accompanying discussion. Figure \ref{fig:sigma_estimator} reports, for each $m$, the empirical averages (dots) and the $5\%$ and $95\%$ quantiles (horizontal lines) of the estimates across the $N$ replications.
The three columns show results for the first diagonal element $(\widehat{\bfSigma}_n)_{11}$ (left), the off-diagonal element $(\widehat{\bfSigma}_n)_{12}$ (center), and the second diagonal element $(\widehat{\bfSigma}_n)_{22}$ (right) of the estimated covariance matrix. Results based on the disjoint-block and sliding-block estimators are shown in blue and yellow, respectively. The solid black lines indicate the true values, obtained from the simulation design described in Section B.1 of the Supplement. 
The rows correspond to different data-generating processes, labeled as models (a)–(c).
For the linear models (a) and (b) the estimates appear nearly unbiased and stable once $m \geq 50$ while for the more complex model (c) a slight bias is visible.
Overall, the sliding-block estimator outperforms the disjoint-block estimator, being less biased in estimating $(\bfSigma_0)_{12}$ and $(\bfSigma_0)_{22}$ and more accurate across all three components.
For model (d), the analytical expression of $R_0(x,y)$ is unknown, and therefore simulation results are not available in this case.
\begin{table}[t!]
{\fontsize{8}{8}\selectfont
\centering
\begin{tabular}{cc|ccc|ccc|ccc|ccc}
\toprule
\multicolumn{2}{c|}{} & \multicolumn{3}{c|}{$\gamma_0$} & \multicolumn{3}{c|}{$a_0(n/k)$} & \multicolumn{3}{c|}{$\bfvartheta_0$} & \multicolumn{3}{c}{$Q_0(1 - 1 / n)$}\\
\midrule
                $n$ & $k$ & 
                BCI & BACI &  FCI &
                BCI & BACI &  FCI &
                BCR & BACR &  FCR &
                BCI & BACI &  FCI  \\
\midrule
\multicolumn{14}{c}{(a) AR(1) model}\\
\midrule
                1000 & 25 & 0.56 & 0.94 & 0.65 
                & 0.51 & 0.81 & 0.84
                & 0.34 & 0.69 & 0.56
                & 0.42 & 0.77 & 0.41  \\
                & 50 & 0.53 & 0.93 & 0.77 
                & 0.53 & 0.88 & 0.89  
                & 0.32 & 0.75 & 0.70 
                & 0.45 & 0.83 & 0.51  \\
                & 100 & 0.51 & 0.91 & 0.83
                & 0.55 & 0.91 & 0.92  
                & 0.30 & 0.80 & 0.79 
                & 0.45 & 0.86 & 0.58  \\
                \hline
                2000 & 50 & 0.52 & 0.94 & 0.79  
                & 0.53 & 0.89 & 0.89
                & 0.31 & 0.76 & 0.71 
                & 0.43 & 0.84 & 0.51  \\
                & 100 & 0.52 & 0.93 & 0.85 
                & 0.54 & 0.91 & 0.91  
                & 0.30 & 0.81 & 0.79 
                & 0.45 & 0.88 & 0.59  \\
                & 200 & 0.50 & 0.92 & 0.88 
                & 0.55 & 0.93 & 0.93  
                & 0.29 & 0.85 & 0.85 
                & 0.45 & 0.89 & 0.65  \\
                \hline
                4000 & 100 & 0.52 & 0.94 & 0.86 
                & 0.55 & 0.92 & 0.92 
                & 0.30 & 0.83 & 0.81
                & 0.46 & 0.89 & 0.60 \\
                & 200 & 0.52 & 0.94 & 0.90
                & 0.54 & 0.93 & 0.93
                & 0.29 & 0.86 & 0.86
                & 0.46 & 0.91 & 0.67 \\
                & 400 & 0.49 & 0.93 & 0.90
                & 0.53 & 0.94 & 0.95
                & 0.28 & 0.89 & 0.89
                & 0.45 & 0.92 & 0.72 \\
\midrule
\multicolumn{14}{c}{(b) ARMA(1,1) model}\\
\midrule
                1000 & 25 & 0.78 & 0.95 & 0.70
                & 0.71 & 0.91 & 0.93
                & 0.52 & 0.81 & 0.58
                & 0.57 & 0.87 & 0.51 \\
                & 50 & 0.65 & 0.97 & 0.76
                & 0.73 & 0.97 & 0.97
                & 0.47 & 0.88 & 0.72
                & 0.55 & 0.91 & 0.61 \\
                & 100 & 0.52 & 0.92 & 0.75
                & 0.60 & 0.97 & 0.97
                & 0.38 & 0.90 & 0.79
                & 0.50 & 0.90 & 0.65 \\
                \hline
                2000 & 50 & 0.69 & 0.99 & 0.81 
                & 0.72 & 0.95 & 0.95 
                & 0.48 & 0.86 & 0.73
                & 0.57 & 0.92 & 0.63 \\
                & 100 & 0.61 & 0.96 & 0.83 
                & 0.71 & 0.98 & 0.98
                & 0.45 & 0.90 & 0.82
                & 0.55 & 0.92 & 0.69 \\
                & 200 & 0.48 & 0.88 & 0.78
                & 0.49 & 0.94 & 0.95
                & 0.31 & 0.90 & 0.83
                & 0.49 & 0.90 & 0.69 \\
                \hline
                4000 & 100 & 0.65 & 0.98 & 0.87 
                & 0.72 & 0.96 & 0.96 
                & 0.45 & 0.89 & 0.82
                & 0.57 & 0.94 & 0.71 \\
                & 200 & 0.57 & 0.95 & 0.87
                & 0.71 & 0.98 & 0.98
                & 0.43 & 0.92 & 0.88
                & 0.54 & 0.93 & 0.74 \\
                & 400 & 0.43 & 0.85 & 0.78
                & 0.34 & 0.86 & 0.89
                & 0.22 & 0.87 & 0.83
                & 0.47 & 0.89 & 0.71 \\
\midrule
\multicolumn{14}{c}{(c) ARCH(1) model}\\
\midrule
                1000 & 25 & 0.90 & 0.97 & 0.78
                & 0.84 & 0.93 & 0.92 
                & 0.77 & 0.87 & 0.69
                & 0.80 & 0.92 & 0.67 \\
                & 50 & 0.87 & 0.95 & 0.82 
                & 0.86 & 0.93 & 0.92
                & 0.74 & 0.87 & 0.76
                & 0.78 & 0.93 & 0.73 \\
                & 100 & 0.82 & 0.90 & 0.81
                & 0.87 & 0.88 & 0.89
                & 0.68 & 0.80 & 0.73
                & 0.76 & 0.91 & 0.75 \\
                \hline
                2000 & 50 & 0.88 & 0.98 & 0.86 
                & 0.85 & 0.95 & 0.94
                & 0.74 & 0.90 & 0.81
                & 0.80 & 0.95 & 0.76 \\
                & 100 & 0.84 & 0.95 & 0.87
                & 0.86 & 0.93 & 0.93
                & 0.71 & 0.89 & 0.83
                & 0.78 & 0.95 & 0.79 \\
                & 200 & 0.79 & 0.90 & 0.84
                & 0.87 & 0.88 & 0.88
                & 0.64 & 0.80 & 0.86
                & 0.76 & 0.91 & 0.79 \\
                \hline
                4000 & 100 & 0.84 & 0.97 & 0.90 
                & 0.86 & 0.94 & 0.94
                & 0.72 & 0.91 & 0.86
                & 0.78 & 0.95 & 0.80 \\
                & 200 & 0.80 & 0.94 & 0.88
                & 0.85 & 0.92 & 0.92
                & 0.67 & 0.88 & 0.86
                & 0.76 & 0.94 & 0.82 \\
                & 400 & 0.74 & 0.88 & 0.83
                & 0.85 & 0.87 & 0.87
                & 0.60 & 0.79 & 0.89
                & 0.74 & 0.90 & 0.80 \\
\midrule
\multicolumn{2}{c|}{} & \multicolumn{3}{c|}{$\gamma_0$} & \multicolumn{3}{c|}{$a_0(n/k)$} & \multicolumn{3}{c|}{$\bfvartheta_0$} & \multicolumn{3}{c}{$Q_{X_{\bar{n} +1} \vert \bfZ_{\bar{n}+1}} (1 - 1 / n)$}\\
\midrule
\multicolumn{14}{c}{(e) ARMA(2,1) model}\\
\midrule
                1000 & 25 & 0.94 & 0.93 & 0.71
                & 0.94 & 0.92 & 0.88 
                & 0.91 & 0.90 & 0.64
                & 0.93 & 0.88 & 0.76 \\
                & 50 & 0.94 & 0.92 & 0.75 
                & 0.95 & 0.93 & 0.90
                & 0.89 & 0.85 & 0.66
                & 0.94 & 0.91 & 0.83 \\
                & 100 & 0.88 & 0.83 & 0.66
                & 0.94 & 0.90 & 0.89
                & 0.73 & 0.65 & 0.49
                & 0.94 & 0.90 & 0.86 \\
                \hline
                2000 & 50 & 0.95 & 0.94 & 0.78 
                & 0.95 & 0.94 & 0.92 
                & 0.91 & 0.90 & 0.73
                & 0.94 & 0.92 & 0.84 \\
                & 100 & 0.93 & 0.91 & 0.77
                & 0.94 & 0.93 & 0.92
                & 0.84 & 0.81 & 0.68
                & 0.94 & 0.92 & 0.89 \\
                & 200 & 0.80 & 0.75 & 0.61
                & 0.92 & 0.90 & 0.89
                & 0.54 & 0.47 & 0.36
                & 0.93 & 0.90 & 0.89 \\
                \hline
                4000 & 100 & 0.94 & 0.93 & 0.82
                & 0.94 & 0.94 & 0.93
                & 0.89 & 0.88 & 0.76
                & 0.94 & 0.93 & 0.89 \\
                & 200 & 0.89 & 0.87 & 0.76
                & 0.94 & 0.93 & 0.92
                & 0.76 & 0.73 & 0.62
                & 0.94 & 0.93 & 0.91 \\
                & 400 & 0.62 & 0.57 & 0.46
                & 0.90 & 0.87 & 0.87
                & 0.25 & 0.22 & 0.16
                & 0.92 & 0.89 & 0.88 \\
\midrule
\multicolumn{14}{c}{(f) AR(1)-GARCH(1,1) model}\\
\midrule
                1000 & 25 & 0.91 & 0.94 & 0.72
                & 0.90 & 0.93 & 0.92
                & 0.85 & 0.89 & 0.64
                & 0.85 & 0.90 & 0.70 \\
                & 50 & 0.91 & 0.96 & 0.75 
                & 0.92 & 0.94 & 0.92
                & 0.79 & 0.86 & 0.66
                & 0.86 & 0.94 & 0.78 \\
                & 100 & 0.78 & 0.84 & 0.62
                & 0.91 & 0.89 & 0.89
                & 0.51 & 0.62 & 0.45
                & 0.83 & 0.92 & 0.79 \\
                \hline
                2000 & 50 & 0.93 & 0.98 & 0.81 
                & 0.93 & 0.96 & 0.95 
                & 0.84 & 0.92 & 0.75
                & 0.88 & 0.96 & 0.81 \\
                & 100 & 0.87 & 0.94 & 0.80
                & 0.93 & 0.94 & 0.94
                & 0.73 & 0.84 & 0.70
                & 0.86 & 0.95 & 0.85 \\
                & 200 & 0.64 & 0.74 & 0.58
                & 0.87 & 0.86 & 0.86
                & 0.31 & 0.46 & 0.36
                & 0.83 & 0.91 & 0.82 \\
                \hline
                4000 & 100 & 0.91 & 0.97 & 0.86
                & 0.93 & 0.96 & 0.95
                & 0.81 & 0.91 & 0.81
                & 0.87 & 0.97 & 0.86 \\
                & 200 & 0.80 & 0.90 & 0.78
                & 0.91 & 0.93 & 0.93
                & 0.60 & 0.77 & 0.67
                & 0.85 & 0.95 & 0.87 \\
                & 400 & 0.44 & 0.56 & 0.45
                & 0.78 & 0.77 & 0.78
                & 0.11 & 0.23 & 0.18
                & 0.77 & 0.88 & 0.81 \\
\bottomrule
            \end{tabular}}
            \caption{Frequentist coverage of $95$\% confidence region and intervals and their Bayesian counterparts for the key quantities.}
\label{table:stat_coverage_regions}
\end{table}

In the second experiment, we simulate $N = 5000$ independent time series from models (a)–(d), each of length $n \in {1000, 2000, 4000}$. For each replication, we compute the frequentist empirical coverage of asymptotic $(1 - \alpha)$ symmetric confidence regions and equi-tailed confidence intervals, along with their Bayesian counterparts. The coverages are obtained using $\tau_I \in {0.9, 0.95, 0.975}$ (with corresponding $k = n(1 - \tau_I)$), and the same values are used in the estimation of $\bfSigma_0$, while $m$ is fixed at 50.
Table \ref{table:stat_coverage_regions} reports the results for the individual parameters $\gamma_0$ and $a_0(n/k)$, the bivariate parameter $\boldsymbol{\vartheta}_0$, and the extreme quantile $Q_0(1 - 1/n)$. The top sections of the table correspond to models (a)–(c). In all three cases, the coverage of the BACR and BACI estimators starts at a reasonable level and converges to the nominal level as the sample size increases. The adjusted regions and intervals substantially outperform their unadjusted counterparts, whose coverage is generally low—often falling below 50\% for the bivariate parameter $\boldsymbol{\vartheta}_0$ and the extreme quantile $Q_0(1 - 1/n)$. Moreover, the adjusted Bayesian credible regions and intervals consistently outperform their frequentist analogues, highlighting the superiority of the Bayesian approach in finite-sample settings.
The results for model (d), reported in Table 1 of Section F.3 of the Supplement, are more nuanced. For exponential marginal distributions, the findings are broadly consistent with those observed for models (a)–(c), although the adjusted Bayesian credible intervals for $\gamma_0$ and $Q_{0} (1 - 1 / n)$ tend to be conservative. In contrast, for power-law marginal distributions, the comparison is less straightforward: unadjusted Bayesian credible intervals provide better coverage of $\gamma_0$ for small and moderate sample sizes; frequentist confidence intervals achieve better coverage of $a_0(n/k)$; and adjusted Bayesian credible intervals deliver superior coverage of $\boldsymbol{\vartheta}_0$ and $Q_0(1 - 1/n)$ at moderate and large sample sizes.

In the third experiment, we apply the dynamic inference scheme described in Section \ref{sec:Dyn_Bayesian}, by simulating $N = 5000$ independent time series from models (e)-(f), each of length $\bar{n}=n + \lceil \sqrt{n} \rceil$, with $v(\bfZ_i)$ and $w(\bfZ_i)$ specified in Example \ref{ex:WARMA}. A $\sqrt{\bar{n}}$-consistent estimator for $v(\bfZ_i)$ is the least squares estimator \citep[][Th. 12.6]{Francq2005}, which we use to then obtain the residuals $(\widehat{X}_1^{(\bar{n})}, \dots, \widehat{X}_n^{(\bar{n})})$, as outlined in \eqref{eq:residuals} and in particular in such an example. The values of $n$, $\tau_I$, and $m$ are chosen as in the second experiment and we used the specific estimator in Corollary E.22 of the Supplement to estimate $\bfSigma_0$. Coverage results for $\gamma_0$, $a_0(n/k)$, $\bfvartheta_0$, and the dynamic extreme quantile $Q_{Y_{\bar{n}+1}\mid\bfZ_{\bar{n}+1}}(1 - 1/n)$ are reported in the bottom rows of Table \ref{table:stat_coverage_regions}.
For model (e), the unadjusted Bayesian credible region and intervals outperform the adjusted versions, as expected, since the innovations $(X_1, \dots, X_n)$ are independent. Coverage levels are close to the nominal values for moderate sample sizes, except for $\bfvartheta_0$. Encouragingly, the coverage of BACR and BACI improves with sample size, gradually approaching that of the unadjusted estimators. Moreover, the Bayesian adjusted credible region and intervals outperform their frequentist counterparts.
For model (f), the BACR and BACI substantially outperform both the unadjusted versions and the frequentist intervals, except in the case of $a_0(n/k)$, where frequentist intervals achieve similar coverage. Notably, the BACI attains coverage above 0.9 for the dynamic extreme quantile $Q_{Y_{\bar{n}+1}\mid\bfZ_{\bar{n}+1}}(1 - 1/n)$ already at moderate sample sizes.

%
\section{Data Analyses}\label{sec:realanalysis}
%

%
\subsection{Short-term interest rates}\label{sed:interests}
%

Models of interest rate dynamics play a central role in bond pricing and in assessing interest rate risk (see, e.g., \citealp{duffie2010}). A common assumption is that interest rates follow a continuous-time diffusion process, observed at discrete intervals (cf. \citealp{cox1985}).
A classical example is the Vasicek model \citep{vasicek1977}, where instantaneous spot rates evolve according to a CARMA(1,0) process (see Example \ref{ex:WARMA}). More recently, \cite{andresen2014} proposed extending this framework to general CARMA($p$, $q$) models to better capture interest rate dynamics. In an empirical study, \cite{thornton2016} evaluated various $(p,q)$ specifications for the Sterling one-month interbank lending rate between 3 January 1978 and 6 November 2008, concluding that a CARMA(2,1) model provides a satisfactory representation of the data.
Note that such a model admits a weak ARMA(2,1) representation for discretely sampled rates, with innovations that are uncorrelated but generally dependent.
\begin{figure}[t!]
	\centering
	\includegraphics[page=1,width=.32\textwidth]{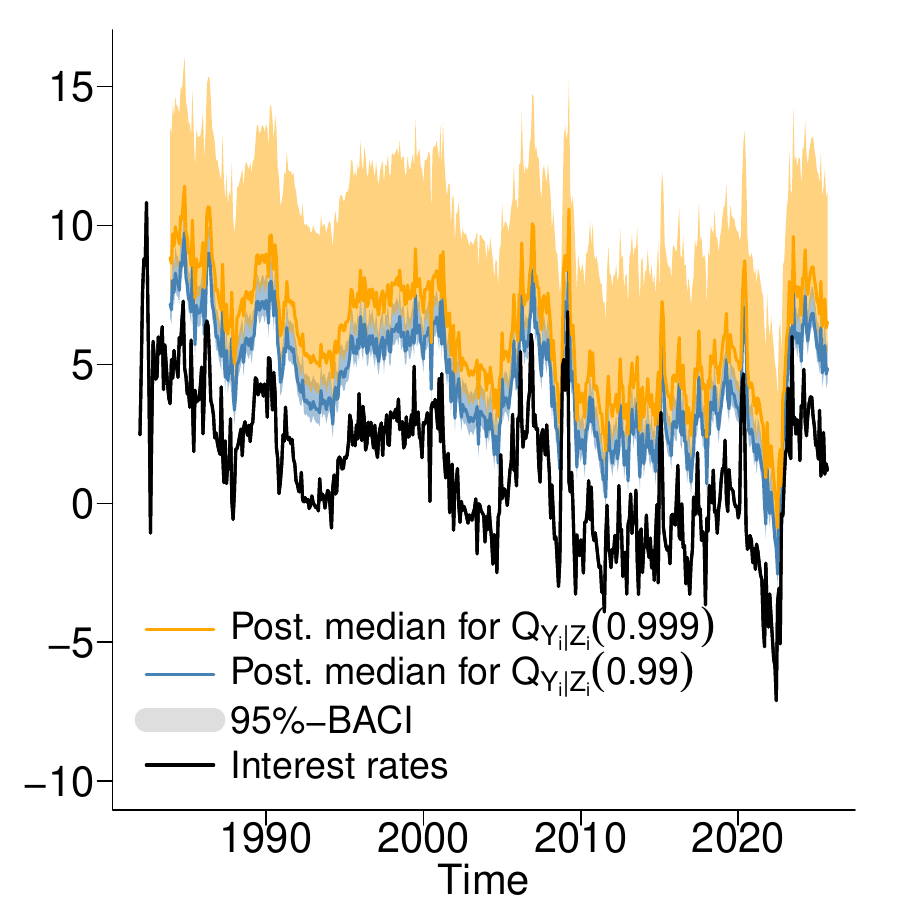}
	\includegraphics[page=2,width=.32\textwidth]{plot_interest_rates.pdf}
	\includegraphics[page=3,width=.32\textwidth]{plot_interest_rates.pdf}
        \caption{US one-month real interest rates,  posterior median and $95$\%-BACI for $Q_{Y_{\bar{n} + 1} \vert \bfZ_{\bar{n}+1}} (\tau_{E})$ for $\tau_{E} = 0.99, 0.999$ (left panel). Posterior density of $Q_{Y_{\bar{n} + 1} \vert \bfZ_{\bar{n}+1}} (\tau_{E})$ with $\tau_{E} = 0.99, 0.995, 0.999$, and corresponding $95$\%-ABCIs for the interest rate at 01 October 2025 (middle panel).
$h$-step ahead inference for $h = 1, \dots, 12$, posterior median and $95$\%-BACI for $Q_{Y_{\bar{n} + h} \vert \bfZ_{\bar{n}+1}} (\tau_{E})$ with $\tau_{E} = 0.99, 0.999$ (right panel).}
\label{fig:real_data_applications}
\end{figure}

From a risk management perspective, accurately assessing both current and future tail risk is of particular importance. In this analysis, we apply our dynamic Bayesian inference scheme introduced in Section \ref{sec:Dyn_Bayesian} to estimate the future $99$\% and $99.9$\% quantiles of the United States one-month real interest rate (REAINTRATREARAT1MO; \citealp[]{fed2025}). The monthly dataset, covering the period from 1 January 1982 to 1 November 2025 and consisting of $\bar{n}=525$ observations, is available through the Federal Reserve Bank of St. Louis data portal (\href{https://fred.stlouisfed.org/}{fred.stlouisfed.org}).

We fit an ARMA(2,1) model with a mean parameter using conditional least squares, obtaining $n = 502$ residuals 
as outlined in Example \ref{ex:WARMA}. The prior for $\gamma$ is specified as $\mathcal{N}(0, 0.4)$, which allocates approximately $80$\% of its mass to the interval $(-1/2, 1/2)$. This reflects a prior belief that one-month interest rates are light- or only mildly heavy-tailed, with finite variance. For $\sigma$, we use the same data-dependent prior used in the simulation study (see Section C.1 of the Supplement for the selection of $k$ and $m$).

The left panel of Figure \ref{fig:real_data_applications} shows the in-sample posterior median and $95$\%-BACIs for $Q_{Y_{\bar{n} + 1} \mid \bfZ_{\bar{n}+1}}(\tau_{E})$ with $\tau_{E} = 0.99, 0.999$, superimposed on the observed time series.
The posterior medians shapes the same trajectory of the data (only shifted aboved), suggesting that the ARMA(2,1) model provides an adequate representation of the mean dynamics, consistent with the findings of \cite{thornton2016}. Approximately $1.0$\% ($0.4$\% and $1.8$\%), and $0.0$\% ($0.0$\% and $0.6$\%) of the in-sample observations exceed the posterior median (and the upper and lower endpoints of the $95$\%-BACI) for $Q_{Y_{\bar{n} + 1} \mid \bfZ_{\bar{n}+1}}(0.99)$ and $Q_{Y_{\bar{n} + 1} \mid \bfZ_{\bar{n}+1}}(0.999)$, respectively. These results indicate that the BACIs provide plausible ranges for future extreme interest rates.
The middle panel of Figure \ref{fig:real_data_applications} displays the posterior densities (obtained from $10^5$ posterior samples by kernel density estimation) of $Q_{Y_{\bar{n} + 1} \mid \bfZ_{\bar{n}+1}}(\tau_{E})$ for $\tau_{E} = 0.99, 0.995, 0.999$ on the out-of-sample date 1 October 2025, along with the corresponding $95$\%-BACIs.
For the less extreme quantile $Q_{Y_{\bar{n} + 1} \mid \bfZ_{\bar{n}+1}}(0.99)$, most of the posterior mass is concentrated within a narrow range around $4.2$\%–$5.6$\% interest rates. In contrast, for higher $\tau_E$ levels, the posterior becomes notably more dispersed, with interest rates above $10$\% still lying within the $95$\%-BACI for $Q_{Y_{\bar{n} + 1} \mid \bfZ_{\bar{n}+1}}(0.999)$.
This highlights both the inherent difficulty of inference on tail events and the importance of reliable uncertainty quantification.

In practice, extrapolations beyond the one-step-ahead horizon are often required.
Naturally, as the forecast horizon increases, the uncertainty surrounding dynamic tail quantiles grows. Simply shifting the posterior distribution of the one-step-ahead tail quantiles by $h$-step-ahead estimates of the conditional mean would fail to capture this additional uncertainty adequately. To address this, we propose the following procedure.
Let $\widehat{v}_{i, h}^{(\bar{n})}$, $i = 1, \ldots, \bar{n} + h$, denote the estimate of the $h$-step-ahead conditional expectation $\mathbb{E}(Y_i \mid \bfZ_{i - h + 1})$.
For ARMA models, such estimates can be obtained iteratively from the corresponding $(h-1)$-step-ahead forecasts.
Then, the posterior distribution of $Q_{Y_{\bar{n} + h} \mid \bfZ_{\bar{n}+1}}(\tau_{E})$ using the estimated $\widehat{v}_{\bar{n} + h, h}^{(\bar{n})}$ and the residuals $\widehat{X}_{i, h}^{(\bar{n})} = Y_{s_n + i} - \widehat{v}_{s_n + i, h}^{(\bar{n})}$, for $i = 1, \ldots, n$, is derived. The estimates $\widehat{v}_{n + h, h}^{(\bar{n})}$ for $h = 1, \ldots, 12$ months, along with the posterior median and $95$\%-BACIs for $Q_{Y_{\bar{n} + h} \mid \bfZ_{\bar{n}+1}}(\tau_{E})$, with $\tau_{E} = 0.99, 0.999$, are displayed by  the right panel of Figure \ref{fig:real_data_applications}.
As expected, both the posterior median and the $95$\%-BACIs increase with the forecast horizon $h$, eventually stabilizing for horizons of five months or more.
This behavior suggests that beyond this point, the past and present provide little additional information beyond the long-run properties of the series.
Notably, while the upper endpoint of the $95$\%-BACI for $Q_{Y_{\bar{n} + 1} \mid \bfZ_{\bar{n}+1}}(0.999)$ remains below $11$\% interest rates, the upper endpoints for horizons $h \geq 4$ already encompass values above $20$\% interest rates.
%
%
\subsection{Electricity demand}\label{sed:eletricity}
%

Electricity utilities and other stakeholders in the power sector require reliable forecasts and proper uncertainty quantification of future demand for operational planning and pricing (see, e.g., \citealp{hong2016} for a recent overview of modeling strategies and applications).
A particularly important measure for reliability planning is the loss of load probability (LOLP) — the probability that electricity demand exceeds the system’s available capacity \citep[]{billinton2008}. The LOLP depends on the joint distribution of available capacity and demand.
A typical target value for the LOLP is $0.1 / 365 \approx 0.027$\%, corresponding to an expectation of one loss-of-load event every ten years \citep[]{billinton1996, meier2006}.
To maintain this target, utility providers must accurately assess the tail behavior of future electricity demand. Specifically, an accurate estimate of the $0.1/365$-quantile of tomorrow’s demand enables the provider to ensure that sufficient generation capacity is available so that demand is unlikely to exceed supply.
Because electricity demand fluctuates substantially throughout the day, the required capacity also varies over time. Moreover, it is well known that demand is influenced by external factors such as meteorological conditions (e.g., temperature) and calendar effects (e.g., local holidays) \citep[]{weron2006}.

In this analysis, we apply our dynamic Bayesian inference procedure to electricity consumption data from the EWZ network (NE5 and NE7)—the primary local electricity provider in the city of Zurich—to forecast future demand between 10:00am and 11:00am, the hour that historically accounts for the largest share of daily peak consumption.
The raw data, covering the period from 1 January 2015 to 21 August 2025 at 15-minute intervals, are publicly available on the Open Data Portal of the city of Zurich (\href{https://data.stadt-zuerich.ch/dataset/ewz_stromabgabe_netzebenen_stadt_zuerich}{data.stadt-zuerich}).
We aggregate the data to daily observations by summing the four 15-minute measurements recorded between 10:00am and 11:00am each day.
To account for weekly seasonality, we compute the difference between each day’s value and that of the corresponding day in the previous week, resulting in a time series of size $3879$, consisting of the weekly-differenced hourly electricity consumption between 10:00 and 11:00 am.
We fit seasonal ARMAX models—incorporating local temperature and calendar effects—using conditional least squares within a rolling five-year window ($\bar{n}=n + \lceil \sqrt{n} \rceil = 1819$), yielding a rolling window of residuals $\widehat{X}_i^{(\bar{n})}$, $i =1,\ldots,n$, with $n = 1776$ .
Since future temperatures are not available in real applications, we mimic the use of forecasts (typically provided by meteorological institutes) by adding independent noise, distributed as $\mathcal{N}(0, 18 / 25\pi)$, to the observed local temperatures. This produces an expected absolute forecast error of approximately $1.2$°C, consistent with observed 24-hour-ahead temperature forecast accuracy (\href{https://content.meteoblue.com/en/research-education/weather-data-accuracy}{meteoblue.com}).
\begin{figure}[t!]
	\centering
        \includegraphics[page=1,width=.32\textwidth]{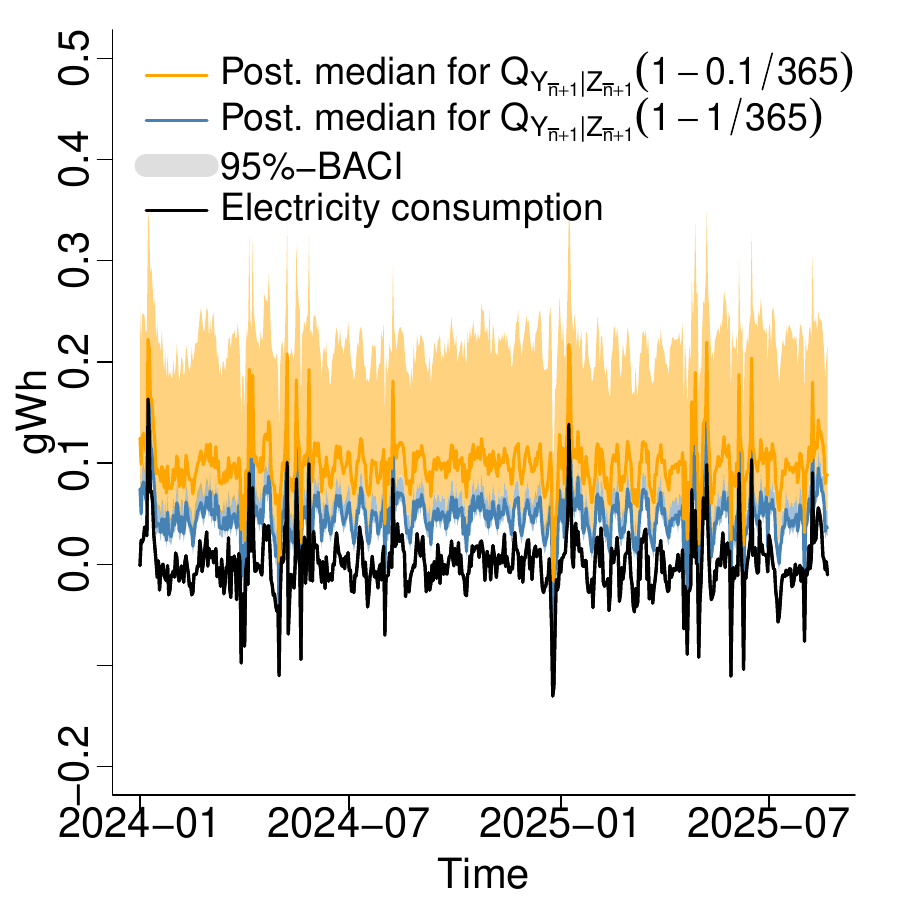}
	\includegraphics[page=2,width=.32\textwidth]{plot_electricity_consumption.pdf}
	\includegraphics[page=3,width=.32\textwidth]{plot_electricity_consumption.pdf}
	\caption{Weekly differenced electricity consumption between 10am to 11am time for the period of 01 January 2024 to 21 August 2025. Posterior median and $95$\%-BACI for $Q_{Y_{\bar{n} + 1} \vert \bfZ_{\bar{n}}} (1 - 1 /  365)$ and $Q_{Y_{\bar{n} + 1} \vert \bfZ_{\bar{n}}} (1 - 0.1 /  365)$ (left panel). Posterior means and BACI of $Q_{Y_{\bar{n} + 1} \vert \bfZ_{\bar{n}}} (1 - 0.1 / 365)$ on 22 August 2025 vs different temperature forecasts (center panel). Posterior density of $Q_{Y_{\bar{n} + 1} \vert \bfZ_{\bar{n}}} (1 - 0.1 / 365)$ on 22 August 2025 for three different temperature forecasts (right panel).}
\label{fig:real_data_applications2}
\end{figure}

Posterior samples for the out-of-sample tail quantiles $Q_{Y_{\bar{n} + 1} \mid \bfZ_{\bar{n}+1}}(\tau_{E})$, with $\tau_{E} = 1 - 1/365$ and $\tau_{E} = 1 - 0.1/365$, are then obtained using the same data-dependent prior as in the interest rate analysis, reflecting a prior belief that electricity consumption exhibits light or only mildly heavy tails.
Further details on model specification and the selection of $k$ and $m$ are provided in section C.2 of the Supplement.
Of the $2060$ out-of-sample observations (approximately $5.64$ years of data), $6$ observations ($5$ and $16$), and $2$ observations ($0$ and $6$), exceed the posterior median (and the upper and lower endpoints of the $95$\%-BACI) of $Q_{Y_{\bar{n} + 1} \mid \bfZ_{\bar{n}+1}}(1 - 1/365)$ and $Q_{Y_{\bar{n} + 1} \mid \bfZ_{\bar{n}+1}}(1 - 0.1/365)$, respectively.
The left panel of Figure \ref{fig:real_data_applications2} shows the posterior medians and $95$\%-BACI from 1 January 2024 onward, plotted at the level of weekly differences for ease of visualization. Note that the posterior medians closely replicate the data trajectory. These results suggest that the model adequately captures the dynamic tail behavior of electricity demand. We then focus on inference for $Q_{Y_{\bar{n} + 1} \mid \bfZ_{\bar{n}+1}}(1 - 0.1/365)$ for 22 August 2025, using the full dataset ($\bar{n}=3869$) to estimate the SARIMA coefficients and the $n = 3816$ residuals.
The center panel of Figure \ref{fig:real_data_applications2} presents the posterior median of $Q_{Y_{\bar{n} + 1} \mid \bfZ_{\bar{n}+1}}(1 - 0.1/365)$ along with the $75$\%-, $90$\%-, and $95$\%-BACIs as a function of the expected temperature in Zurich between 10:00am and 11:00am on that date.
Although temperature clearly influences the extreme quantile, its effect is overshadowed by the overall uncertainty in $Q_{Y_{\bar{n} + 1} \mid \bfZ_{\bar{n}+1}}(1 - 0.1/365)$.
To illustrate this further, the right panel of Figure \ref{fig:real_data_applications2} displays the corresponding posterior densities for three different temperature forecasts (obtained from a posterior sample of size $10^5$ and using kernel density estimation).
While relatively low consumption levels around $0.43$ GWh remain plausible when the expected temperature is $20$°C, they become increasingly unlikely on very hot days with temperatures around $30$°C. Conversely, higher consumption levels above $0.50$–$0.55$ GWh are plausible within a 10-year horizon under all temperature scenarios.

\bibliographystyle{chicago} 
\bibliography{references_sr}

\end{document}